\documentclass[a4paper,12pt]{article}
\usepackage[utf8]{inputenc}
\usepackage[T1]{fontenc}
\usepackage{microtype}
\usepackage{amssymb,latexsym}
\usepackage[hidelinks]{hyperref}
\usepackage{mathtools}
\usepackage{todonotes}
\usepackage{cite}
\usepackage{enumitem}
\usepackage{cleveref}
\usepackage{float}
\usepackage{booktabs}
\usepackage{siunitx}

\makeatletter
 \g@addto@macro\@verbatim\small
\makeatother

\long\def\new#1\endnew{{\bf #1}}
\long\def\del#1\enddel{}

\let\0=\over      \let\1=\vec      \def\2{{1\over2}}    \let\3=\ss
\let\4=\underline \let\5=\overline \let\6=\partial      \def\7#1{{#1}\llap{/}}
\def\8#1{{\textstyle{#1}}}         \def\9#1{{\ifmmode{\pmb{#1}}\else\bf#1\fi}}

          \def\({\left(}       \def\){\right)}

\def\eeql#1 {\label{#1}\eeq}      \let\nn=\nonumber
\def\beq{\begin{equation}}      \def\eeq{\end{equation}}
\def\bea{\begin{eqnarray}}      \def\eea{\end{eqnarray}}

\let\and=\wedge                   
      \let\iff=\Leftrightarrow
                
\let\bra=\langle        \let\ket=\rangle        \def\<#1\>{\bra #1 \ket}

\def\rel#1 #2{\buildrel #1 \over {#2}}
\def\fnote#1#2{\begingroup\def\thefootnote{#1}\footnote{#2}
                \addtocounter{footnote}{-1}\endgroup}

          \let\d=\delta   
         \let\th=\theta

             \let\D=\Delta

\def\IR{{\mathbb R}}  
  
\def\IZ{{\mathbb Z}}

\crefname{figure}{Fig.}{Figs.}
\crefname{table}{Table}{Tables}

\begin{document}

\begin{center}
{\Huge\bf  All Weight Systems for \\[3mm] Calabi--Yau Fourfolds \\[6mm] from Reflexive Polyhedra}\vskip 15mm
Friedrich Schöller\fnote{*}{e-mail: schoeller@hep.itp.tuwien.ac.at} and \,
Harald Skarke\fnote{$\dagger$}{e-mail: skarke@hep.itp.tuwien.ac.at}\\[3mm]
Institut für Theoretische Physik, Technische Universität Wien\\
Wiedner Hauptstraße 8--10, 1040 Wien, Austria

\vfill                  {\bf ABSTRACT }
\end{center}
For any given dimension $d$, all reflexive $d$-polytopes can be found (in principle) as subpolytopes of a number of maximal polyhedra that are defined in terms of $(d+1)$-tuples of integers (weights), or combinations of $k$-tuples of weights with $k<d+1$.
We present the results of a complete classification of sextuples of weights
pertaining to the construction of all reflexive polytopes in five dimensions.
We find \num{322383760930} such weight systems. \num{185269499015} of them give rise directly to reflexive polytopes and thereby to mirror pairs of Calabi--Yau fourfolds.
These lead to \num{532600483} distinct sets of Hodge numbers.
\vfill 
\thispagestyle{empty}\newpage
\pagestyle{plain}

\setcounter{page}{1}

\section{Introduction}
When the relevance of Calabi--Yau manifolds to string compactification was first recognized \cite{Candelas:1985en}, only very few such manifolds were known and it was hoped that a direct enumeration of all possibilities might lead to the identification of the string vacuum describing our universe.
This hope has not been fulfilled.
On the one hand there is still no general algorithm for a complete
classification of all possible topologies of Calabi--Yau manifolds, and on
the other hand it is by now understood that there are so many different
constructions involving bundles, fluxes, D-branes etc.\ that probably even a
complete list of all geometries would not get us very far.

Nevertheless it is important to have large lists of Calabi--Yau manifolds,
both to scan for the possibility of finding standard model like physics, and
to have a playground for doing statistics and checking hypotheses.
Typically such lists are the results of solving classification problems for
specific types of Calabi--Yau manifolds.
The first computer aided classification was that of complete intersection
Calabi--Yau threefolds \cite{Candelas:1987kf}, which resulted in
250 distinct pairs of Hodge numbers $(h^{1,1},h^{1,2})$.
A significantly larger list consists of hypersurfaces in weighted projective
spaces, subject to the condition that the hypersurfaces feature no
singularities beyond the ones already present in the ambient spaces.
There are \num{7555} models of this type \cite{Klemm:1992bx,hep-th/9205004},
and together with \num{3284} similar models of Landau--Ginzburg type they
lead to \num{2997} Hodge number pairs; including all abelian orbifolds
of these models gives rise to 800 further pairs \cite{Kreuzer:1992iq}.

The richest source of models up to now has been toric geometry,
a branch of algebraic geometry that allows for reasonably simple and explicit
characterizations of algebraic varieties of dimension $d$ in terms of
elementary data pertaining to lattices of the type $\IZ^d$.
Calabi--Yau hypersurfaces in toric varieties can be described
via reflexive polytopes \cite{alg-geom/9310003}.
A generalization via reflexive Gorenstein cones \cite{alg-geom/9402002}
provides data for gauged linear sigma models
\cite{hep-th/9301042} that may correspond to higher codimension submanifolds
in toric varieties, Landau--Ginzburg models or hybrids.

Reflexive polytopes of dimension $d=4$ account for the largest known list of
Calabi--Yau threefolds.
Their classification proceeded in two steps.
First a set $S$ of ``maximal'' polytopes was constructed, with the
property that any reflexive polytope would have to be a subpolytope of one of
the elements of $S$.
Then the remaining (straightforward in principle but very tedious in
practice) task was to find all subpolytopes of the elements of $S$.
Any such maximal polytope can be described with the help of one or more
weight systems (collections of positive numbers).
The weight systems pertaining to reflexive 4--polytopes
were found in Ref.~\cite{alg-geom/9603007}.
There it was also shown that each of them gives rise to a maximal polytope
that is actually reflexive, which is a property that need not hold for $d>4$.
Alternatively one can think of the weight system as defining a weighted
projective space that can be partially
resolved into a toric variety that contains a smooth Calabi--Yau hypersurface.
This first step alone resulted already in \num{184026} models and \num{10237} sets of
Hodge data.
After an intermediate step of combining lower-dimensional weight systems into
\num{17320} further maximal polytopes \cite{hep-th/9703003} and several processor
years of searching for subpolytopes, a list of \num{473800776} reflexive polytopes
emerged \cite{hep-th/0002240}; these gave rise to \num{30108} Hodge number pairs.

It should be possible to find many further Calabi--Yau threefolds by
constructing reflexive Gorenstein cones.
For this problem only the first step of classifying the pertinent weight systems
has been taken \cite{Skarke:2012zg}.

With the advent of F--theory \cite{hep-th/9602022} Calabi--Yau fourfolds
(at least elliptically fibred ones) acquired phenomenological relevance.
The strategies for constructing threefolds can be applied to the fourfold case
as well.
Soon there existed the complete list of \num{1100055} hypersurfaces in weighted
projective spaces \cite{hep-th/9812195}
which gave rise to \num{667954} triples of numbers $(h^{1,1},h^{1,2},h^{1,3})$
(the only other nontrivial Hodge number $h^{2,2}$ can be computed from these).
More recently the (smaller) list of complete intersection fourfolds was also
found \cite{arXiv:1303.1832}.

Clearly we would expect much larger numbers of fourfolds from reflexive polytopes.
An amusing way of obtaining a rough guess of the order of magnitude is by taking a look at Table \ref{nRP}.
\begin{table}
\centering
\begin{tabular}{@{}rrrrr@{}}
  \toprule
  $d$ & $n_\mathrm{RP}$ & $\mathrm{ld}(4+\mathrm{ld}(n_\mathrm{RP}))-1$ \\
  \midrule
  1 & 1 & 1 \\
  2 & 16 & 2 \\
  3 & 4319 & 3.0069 \\
  4 & \num{473800776} & 4.0365 \\
  \bottomrule
\end{tabular}
\caption{Numbers $n_\mathrm{RP}$ of reflexive polytopes of dimension $d$ and a particular function of $n_\mathrm{RP}$ that involves two dual logarithms.}
\label{nRP}
\end{table}
This table gives the number $n_\mathrm{RP}(d)$ of reflexive polytopes for every dimension $d$ for which it is known, as well as the function
\beq f(n_\mathrm{RP})=\mathrm{ld}(4+\mathrm{ld}(n_\mathrm{RP}))-1 \eeq
of $n_\mathrm{RP}(d)$ which involves twice taking a logarithm with respect to base 2.
As Table \ref{nRP} shows, $f(n_\mathrm{RP}(d))\ge d$,
with equality for $d=1,2$ and small deviations for $d=3,4$.
Upon inverting this to $n_\mathrm{RP}(d)\ge 2^{2^{d+1}-4}$ and assuming a similar relationship for $d=5$, we would estimate $n_\mathrm{RP}(5)$
to exceed $2^{2^6-4}\approx 1.15\times 10^{18}$.
Since we lack the capacity to store more than a million TRPs (tera-reflexive-polytopes) we decided to aim for a more moderate goal.

The natural thing to do is, of course, to look for the weight systems
corresponding to $d=5$.
In the present paper we describe how we did that and what results we obtained.
Section \ref{CY2w} contains a brief summary of the required concepts as well as an outline of the classification algorithm for reflexive polytopes.
Section \ref{alg} describes our algorithm for finding the weight systems, which is an
improved version of the one used in Refs.~\cite{alg-geom/9603007,Skarke:2012zg}.
Section \ref{ill} provides an illustration of the algorithm and section \ref{imp} is concerned with its implementation.
In section \ref{res} we present and discuss our results.
An appendix contains a number of plots and diagrams that should make some of
the rich structure of our data visible.

\section{From Calabi--Yau manifolds to weights}\label{CY2w}

\subsection{General aspects}

Toric geometry is usually formulated with reference to a dual pair of lattices $M\simeq\IZ^d$, $N=\mathrm{Hom}(M,\IZ)$ and their real extensions $M_\IR\simeq\IR^d$, $N_\IR\simeq\IR^d$.
A \emph{lattice polytope} $\D\subset M_\IR$ is a polytope, i.e.\ the convex hull of a finite number of points, with vertices in $M$.
We use the words polytope and polyhedron interchangeably.
Following \cite{hep-th/9805190} we say that a polytope has the \emph{IP property} (or, is an ``IP polytope'') if the origin is in its interior.
The \emph{dual} of a set $\D\subset M_\IR$ is
\beq \D^* = \{y\in N_\IR: \<y,x\>+1\ge 0 ~~\forall x\in\D\};\eeq
the dual of an IP polytope $\D$ is itself an IP polytope with $\D^{**} = \D$.
An IP polytope $\D$ is called \emph{reflexive} if both $\D$ and $\D^*$ are lattice polytopes.

Batyrev \cite{alg-geom/9310003} realized that mirror pairs of Calabi--Yau manifolds can be described via reflexive polytopes.
The fan (i.e., set of cones) over some triangulation of the surface of $\D^*\subset N_\IR$ provides the data for a toric variety, and the lattice points of $\D\subset M_\IR$ correspond to the monomials occurring in the polynomial describing a Calabi--Yau hypersurface in that variety.
In this context mirror symmetry just corresponds to swapping $\D$ and $\D^*$.
This symmetry manifests itself, in particular, by an exchange $h^{i,j}\leftrightarrow h^{i,d-1-j}$ of the Hodge numbers of the corresponding $(d-1)$--dimensional Calabi--Yau manifolds.
The following formula summarizes results of Batyrev \cite{alg-geom/9310003} and Batyrev and Dais \cite{alg-geom/9410001} for Hodge numbers of the type $h^{1,i}$:
\begin{align}
  h^{1,i}&=\d_{1i}\(l(\D^*)-d-1-\sum_{\mathclap{\mathrm{codim}\,\th^* =1}}l_\mathrm{int}(\th^*)\)\nn\\
        &\quad+ \d_{d-2,i}\(l(\D)-d-1-\sum_{\mathclap{\mathrm{codim}\,\th =1}}l_\mathrm{int}(\th)\)\nn\\
        &\quad+ \sum_{\mathclap{\mathrm{codim}\,\th^* =i+1}}l_\mathrm{int}(\th^*)l_\mathrm{int}(\th). \label{baho}
\end{align}
Here $l$ gives the number of lattice points of some polytope and $l_\mathrm{int}$ the number of interior lattice points;
$\th$ and $\th^*$ denote mutually dual faces of $\D$ and $\D^*$, respectively,
with codimensions as indicated under the summation symbols.
For the present case of Calabi--Yau fourfolds (i.e.\ $d=5$) the only further non-trivial Hodge number is $h^{2,2}$ which depends on the others via the well known relation
\beq h^{2,2}=44+4h^{1,1}-2h^{1,2}+4h^{1,3}. \eeql{h22formula}

\subsection{Classification of reflexive polytopes}
The main idea of the classification algorithm of \cite{hep-th/9512204,math/0001106} is to look for a set $S=\{\D_1, \D_2,\ldots \}$ of lattice polytopes such that any reflexive polytope $\D$ is contained in at least one of the $\D_i$.
Since duality inverts subset relations, $\D\subseteq\tilde\D \Longleftrightarrow \D^* \supseteq \tilde\D^*$, every reflexive polytope must then contain at least one of $\D_1^*, \D_2^*,\ldots$.

This motivates the definition of a \emph{minimal polytope} $\nabla\subset N_\IR$ as a polytope that has the IP property, whereas the convex hull of any proper subset of the set of its vertices fails to have it.

Properties of minimal polytopes were analysed in \cite{hep-th/9512204}.
It turns out that a minimal polytope is either a simplex with the origin in the interior (``{IP simplex}'') or the convex hull of a number of lower-dimensional simplices of that type; for any given dimension there is a finite number of combinatorial ways in which a minimal polytope can consist of several lower-dimensional IP simplices.
To any IP simplex of dimension $d$ with vertices $V_{1}, \dots, V_{d+1}$ we can assign a \emph{weight system} (array of weights) ${\mathbf q} \subset \IR^{d+1}_{>0}$ via $\sum_i q_i V_i=0$.
The definition of ${\mathbf q}$ is unique up to rescaling.
In the case where the $V_i$ are lattice points we can use this freedom to make the $q_i$ integer; alternatively we can use a
convention such as $\sum_i q_i=1$.
Minimal polytopes consisting of more than one IP simplex are described by
\emph{combined weight systems} (matrices of weights).

Given a minimal polytope $\nabla$ constructed with a (combined) weight system,
its dual $\nabla^*$ will usually not be a lattice polytope.
In order to be relevant for our classification problem, $\nabla^*$
must however contain a lattice polytope with the IP property.
This statement only makes sense once we know to which pair of lattices we
are referring.
The coarsest lattice 
for which $\nabla$ is a lattice polytope 
is just the lattice linearly generated by the vertices of $\nabla$;
this lattice $N_\mathrm{coarsest}$, which is determined by the (combined) weight system, must be a sublattice of any other lattice $N$ that contains the vertices of $\nabla$.
Then $N_\mathrm{coarsest}\subseteq N$ implies $M\subseteq M_\mathrm{finest}$ (the lattice dual to $N_\mathrm{coarsest}$), so the convex hull conv($\nabla^*\cap M$) can have the IP property only if conv($\nabla^*\cap M_\mathrm{finest}$) has it.

We say that a (combined) weight system has the IP property if conv($\nabla^*\cap M_\mathrm{finest}$) has it, where $\nabla$ is the corresponding minimal polytope.
One can easily show \cite{math/0001106} that, for a combined weight system to have the IP property, it is necessary that every single weight system occurring in it has this property; we shall refer to such weight systems as IP weight systems.

In Ref.~\cite{alg-geom/9603007} the IP weight systems for $d\le 4$ were found: there are 3/95/\num{184026} for $d$ equal to 2/3/4, respectively.
In addition there are 1/21/\num{17320} combined weight systems giving rise to non-simplicial 2/3/4-dimensional minimal polytopes with the IP property \cite{hep-th/9703003}.
The remaining task in the classification \cite{hep-th/9805190,hep-th/0002240} of all reflexive polytopes of dimension up to 4 was to find all reflexive subpolytopes (both on $M_\mathrm{finest}$ and on its sublattices) of these  4/116/\num{201346} polytopes and to ensure that every isomorphism class of polytopes was counted only once; this was achieved by the introduction of a suitable normal form for reflexive polytopes.

In the present work we report how we found all weight systems with the IP property for $d=5$.

\section{Algorithm}\label{alg}
Given a weight system $\mathbf{q}$ we need an efficient description of the polytope
determined by $\mathbf{q}$.
This is achieved as follows \cite{hep-th/9512204,math/0001106}.
If $V_1, \ldots , V_n$ are the vertices of $\nabla$ satisfying $\sum_i q_i V_i=0$,
we define an embedding map for $\nabla^*\subset M_\IR$ via
\beq M_\IR\to \IR^n, ~~~ X\mapsto{\mathbf y}=(y_1,\ldots,y_n) \hbox{ with }
y_i=\<V_i,X\>. \eeq
Under this map the image of $M_\IR$ is  the linear subspace of $\IR^n$ for which $\sum_i q_i y_i=0$.
The image of $\nabla^*$ also satisfies $y_i\ge -1$ for all $i$, and the image of $M_\mathrm{finest}$ is $\IZ^n\cap \{\mathbf{y}:\sum_i q_i y_i=0\}$.
Here we have $n=d+1$, but the same construction works for a combination of $k$ weight systems and $n=d+k$.

Clearly $\mathbf q$ has the IP property if and only if $\mathbf{0}$ is in the interior of the convex hull of
\beq \IZ^n\cap \{\mathbf{y}:\sum_i q_i y_i=0, y_i\ge -1 ~\forall i\}.\eeql{Dq}
Upon passing to new coordinates $x_i = y_i + 1$ (and thereby turning our linear subspace into an affine one) the condition $\sum_i q_i y_i=0$ changes to $\sum_i x_i q_i=r$ whereby the normalization $r=\sum_i q_i$ becomes relevant.
For $r=1$ we can restate the IP condition as $(1,\ldots,1) \in \mathrm{int}(\D_{\mathbf q})$ with
\beq
\D_{\mathbf q}=
\mathrm{conv}(\{(x_1,\ldots,x_n):x_i \in \IZ_{\ge 0}, \sum_i x_i q_i=1\}).
\eeql{ipws}
This can hold only if all weights obey $q_i\le 1/2$ (if $q_i>1/2$ then $x_i\in\{0,1\}$ for all $\mathbf{x} \in \D_{\mathbf q}$, leading to a violation of the IP condition).
For $n>2$ at most one of the $q_i$ can be equal to $1/2$ (otherwise $\sum_i q_i>1$).
Furthermore it is not difficult to see that for $q_n=1/2$ the IP condition amounts to $(2,\ldots,2) \in \mathrm{int}(\D_{(q_1,\ldots, q_{n-1})})$.
This allows us to restrict our attention to weights smaller than $1/2$, with a convenient split of our problem into finding $n=6$ weights with $r=1$ and $(1,1,1,1,1,1)\in\mathrm{int}(\D_{\mathbf q})$ or $n=5$ weights with $r=1/2$ and $(2,2,2,2,2)\in\mathrm{int}(\D_{\mathbf q})$, respectively.

Any set of $n$ linearly independent $\mathbf x^{(i)}\in \D_{\mathbf q}$ will determine $\mathbf q$.
We use this fact for the classification, starting with
${\mathbf x}^{(0)}=(1,\ldots,1)$ or $(2,\ldots,2)$ and continuing by successively adding further lattice points ${\mathbf x}^{(i)}$, thereby restricting the set of allowed $\mathbf q$'s.
Having chosen ${\mathbf x}^{(0)},\ldots, {\mathbf x}^{(k)}$ we can pick an arbitrary $\mathbf{ \tilde q}$ such that $\mathbf x^{(j)}\cdot\mathbf{\tilde q}=1$
(with $\mathbf x \cdot \mathbf q := \sum_i x_i q_i$)
for all $j=0,\ldots,k$;
if $\mathbf{\tilde q}$ has the IP property we add it to our list.
If  $k+1=n$ then $\mathbf{ \tilde q}$ is the unique weight system compatible with
${\mathbf x}^{(0)},\ldots, {\mathbf x}^{(k)}$ and we are finished with this branch
of the construction.
Otherwise, note that ${\mathbf x}^{(0)}$ (which satisfies $\mathbf x^{(0)}\cdot \mathbf{\tilde q}=1$) cannot be interior to $\D_{\mathbf q}$ for $\mathbf q \ne\mathbf{\tilde q}$ unless 
$\D_{\mathbf q}$ contains points satisfying $\mathbf x\cdot\mathbf{\tilde q}<1$.
Therefore it suffices to consider each of the finitely many lattice points obeying $x_i\ge 0$ for all $i$ and $\mathbf x\cdot\mathbf{\tilde q}<1$ as the next chosen point ${\mathbf x}^{(k+1)}$.
Given the finite number of choices at each branching level $k\in\{0,\ldots, n-2\}$ we are
bound to eventually find all allowed weight systems.

As in \cite{Skarke:2012zg} we use the following method for finding suitable
$\mathbf{\tilde  q}$'s.
Every set of linearly independent ${\mathbf x}^{(0)},\ldots, {\mathbf x}^{(k)}$
determines the $(n-k-1)$--dimensional polytope
\beq\{\mathbf{q}:q_i\ge 0,~ \mathbf x^{(j)}\cdot\mathbf q=1 ~\forall j=0,\ldots,k\}\eeql{qpol}
in $\mathbf q$--space.
The vertices of this polytope can be computed efficiently by using the
$(n-k)$--dimensional polytope of the previous step.
We simply take $\mathbf{\tilde  q}^{(k)}$ as the average of these vertices.

The algorithm as presented so far has the disadvantage of finding weight
systems many times, both in terms of identical copies
${\mathbf q'} = {\mathbf q}$ and of permutation equivalent ones,
$q_i' = q_{\pi(i)}$, where $\pi$ represents an element of the group $S_n$ of
permutations of the coordinates.

The following strategy gets rid of identical copies almost completely.
Essentially, if we want to find a given weight system $\mathbf q$ precisely
once, there must be a unique sequence of ${\mathbf x}^{(i)}$ resulting in
$\mathbf q$.
Such a sequence can be defined by ${\mathbf x}^{(k+1)}$ being the lattice
point in $\D_{\mathbf q}$ that minimizes $\mathbf x \cdot\mathbf{\tilde q}^{(k)}$;
if the minimum value occurs for more than one lattice point,
${\mathbf x}^{(k+1)}$ is taken as the lexicographically largest one among them
(this is equivalent to a very small deformation of $\mathbf {\tilde q}^{(k)}$).
This results in a unique sequence of lattice points
${\mathbf x}^{(0)},{\mathbf x}^{(1)},\ldots ,{\mathbf x}^{(n-1)}$
defining $\mathbf q$ except in those rare cases in which
$\mathbf q = \mathbf {\tilde q}^{(k)}$ for some $k<n-1$.
During the execution of the algorithm $\mathbf q$ is of course not yet known.
The conditions defined above are implemented as follows: given
${\mathbf x}^{(0)},\ldots ,{\mathbf x}^{(k)}$
we determine the set of all nonnegative lattice points in the affine space
spanned by them and abandon this branch of the
recursion unless all of ${\mathbf x}^{(0)},\ldots ,{\mathbf x}^{(k)}$
satisfy the above criteria within that space.

Redundancies from permutation equivalences can be reduced by keeping track
of which subgroup $G^{(k)}$ of the original group $G^{(0)}=S_n$ of coordinate
permutations leaves each of the points
${\mathbf x}^{(0)},\ldots ,{\mathbf x}^{(k)}$ invariant;
then one proceeds with a given ${\mathbf x}^{(k+1)}$ only if it is the
lexicographically largest one within its $G^{(k)}$-orbit.

Very explicitly, having computed $\mathbf{\tilde q}^{(k-1)}$ our algorithm
determines all points $\mathbf x^{(k)} \in \IZ_{\ge 0}^n$ with
$\mathbf{x}^{(k)} \cdot \mathbf{\tilde q}^{(k-1)} < 1$ and rejects the new
point $\mathbf{x}^{(k)}$ if any of the following conditions, which are checked in the given order, holds:
\begin{enumerate}
\item\label{item:1} $x^{(k)}_i \le 1/r$ for all $i$,
\item\label{item:2} $\sum_i x^{(k)}_i \le 2$, 
\item\label{item:3} $\mathbf{x}^{(k)}$ is not the lexicographically largest after application of allowed coordinate permutations,
\item\label{item:4}
  \begin{enumerate}
  \item $\mathbf{x}^{(k)} \cdot \mathbf{\tilde q}^{(j)} < \mathbf{x}^{(j+1)} \cdot \mathbf{\tilde q}^{(j)}$ or
  \item $\mathbf{x}^{(k)} \cdot \mathbf{\tilde q}^{(j)} = \mathbf{x}^{(j+1)} \cdot \mathbf{\tilde q}^{(j)}$ and $\mathbf{x}^{(k)} > \mathbf{x}^{(j+1)}$
  \end{enumerate}
  for some $j < k-1$,
\item\label{item:5} for some $j < k$ there exists a point $\mathbf{x}$ with $x_i \ge 0$ that lies on the line through $\mathbf{x}^{(k)}$ and $\mathbf{x}^{(j)}$ but not between $\mathbf{x}^{(k)}$ and $\mathbf{x}^{(j)}$,
\item\label{item:6} the sequence $\mathbf{x}^{(0)}, \dots, \mathbf{x}^{(k)}$ does not allow a positive weight system,
\item\label{item:7} the set of nonnegative lattice points in the affine span of $\{{\mathbf x}^{(0)},\ldots ,{\mathbf x}^{(k)}\}$
  (which is computed as the set of $\mathbf x \in \D_{\mathbf{\tilde q}^{(k)}}$ that have an inner product of 1 with every vertex of the $\mathbf q$--space polytope (\ref{qpol})) contains an $\mathbf x$ with
  \begin{enumerate}
  \item $\mathbf{x} \cdot \mathbf{\tilde q}^{(j)} < \mathbf{x}^{(j+1)} \cdot \mathbf{\tilde q}^{(j)}$ or
  \item $\mathbf{x} \cdot \mathbf{\tilde q}^{(j)} = \mathbf{x}^{(j+1)} \cdot \mathbf{\tilde q}^{(j)}$ and $\mathbf{x} > \mathbf{x}^{(j+1)}$
  \end{enumerate}
  for some $j < k$.
\end{enumerate}
These checks are not independent: items 1 and 2 each imply number 6 and items 4 and 5 each imply number 7.
The earlier checks are included because they are so much simpler than the later ones that they result in a reduction of computation time.

\section{Illustration of the algorithm}\label{ill}

In this section we will demonstrate explicitly how our algorithm works by following a specific path in the recursive tree for $n = 5$, $r = 1/2$ from the root at ${\mathbf x}^{(0)}=(2,2,2,2,2)$ to its tip.
We will explain at each branch point some of the considerations that play a role, with references to specific items in the list at the end of the last section.

We start with the subset of $\mathbf q$-space defined by $q_i \ge 0$.
The condition $\sum_i q_i = r = 1/2$, which is equivalent to
$\mathbf x^{(0)} \cdot \mathbf q = 1$, determines a simplex in $\mathbf q$-space which we describe by the matrix
\begin{align}
  \begin{pmatrix}
    1/2 & 0 & 0 & 0 & 0 \\
    0 & 1/2 & 0 & 0 & 0 \\
    0 & 0 & 1/2 & 0 & 0 \\
    0 & 0 & 0 & 1/2 & 0 \\
    0 & 0 & 0 & 0 & 1/2
  \end{pmatrix};
\end{align}
here and later we encode a $\mathbf q$-space polytope by a matrix whose lines correspond to the vertices.
The average of the vertices gives the first candidate for an IP weight system:
\begin{align}
  \mathbf{\tilde q}^{(0)} = (1/10, 1/10, 1/10, 1/10, 1/10).
\end{align}

\paragraph{Point 1}\hfill\\
The point $\mathbf x^{(1)}$ must satisfy $x^{(1)}_i \in \IZ_{\ge 0}$ and $\mathbf x^{(1)} \cdot\mathbf{\tilde q}^{(0)} < 1$, i.e., $\sum_i x^{(1)}_i < 10$.
Points $\mathbf x$ with $x_i \le 1/r = 2$ for all $i$ do not lead to positive weight systems so they are excluded (cf.\ \cref{item:1}).
There are 1760 points that satisfy the conditions.
Taking only the lexicographically largest ones from orbits of coordinate permutations (cf.\ \cref{item:3}) reduces the number to 63 points: $(9, 0, 0, 0, 0)$, $(8, 1, 0, 0, 0)$, $(8, 0, 0, 0, 0)$, $(7, 2, 0, 0, 0)$, $(7, 1, 1, 0, 0)$, \dots, $(3, 0, 0, 0, 0)$. Furthermore, the points $(3, 1, 1, 1, 1)$, $(3, 2, 1, 1, 1)$, $(3, 2, 2, 1, 1)$, $(3, 3, 1, 1, 1)$, $(4, 1, 1, 1, 1)$, $(4, 2, 1, 1, 1)$, and $(5, 1, 1, 1, 1)$ all are excluded because they lie on lines between $(2, 2, 2, 2, 2)$ and another allowed point (cf.\ \cref{item:5}), thereby violating minimality of $\mathbf x^{(1)} \cdot\mathbf{\tilde q}^{(0)}$. This leads to 56 allowed points.
For this example we pick the point
\begin{align}
  \mathbf x^{(1)} = (3, 0, 0, 0, 0).
\end{align}
The condition $\mathbf x^{(1)} \cdot \mathbf q = 1$ restricts the allowed region in $\mathbf{q}$-space to the simplex
\begin{align}
  \begin{pmatrix}
    1/3 & 1/6 & 0 & 0 & 0 \\
    1/3 & 0 & 1/6 & 0 & 0 \\
    1/3 & 0 & 0 & 1/6 & 0 \\
    1/3 & 0 & 0 & 0 & 1/6
  \end{pmatrix},
\end{align}
and we find another candidate for an IP weight system:
\begin{align}
  \mathbf{\tilde q}^{(1)} = (1/3, 1/24, 1/24, 1/24, 1/24).
\end{align}

\paragraph{Point 2}\hfill\\
At this stage, up to coordinate permutations, there are 1164 choices for the point $\mathbf x^{(2)}$ that lead to positive weight systems.
Some of those are excluded by demanding that $\mathbf x^{(k+1)} \cdot\mathbf{\tilde q}^{(k)}$ is minimal among all points leading to the same final $\mathbf q^{(4)}$ (cf.\ \cref{item:7}).
For example, the choice $\mathbf x^{(2)} = (0, 7, 3, 3, 3)$ is not allowed because $\mathbf x \cdot\mathbf{\tilde q}^{(1)} < \mathbf x^{(2)} \cdot\mathbf{\tilde q}^{(1)}$
for $\mathbf x = \mathbf x^{(0)} + 2 \left( \mathbf x^{(1)} - \mathbf x^{(0)} \right) + 2 \left( \mathbf x^{(2)} - \mathbf x^{(0)} \right) = (0, 8, 0, 0, 0)$.
This leaves us with 803 candidates, one of them being
\begin{align}
  \mathbf x^{(2)} = (0, 7, 4, 0, 0).
\end{align}
This point leads to the polytope
\begin{align}
  \begin{pmatrix}
    1/3 & 1/9 & 1/18 & 0 & 0 \\
    1/3 & 1/7 & 0 & 1/42 & 0 \\
    1/3 & 1/7 & 0 & 0 & 1/42
  \end{pmatrix}
\end{align}
in $\mathbf q$-space and the weight system
\begin{align}
  \mathbf{\tilde q}^{(2)} = (1/3, 25/189, 1/54, 1/126, 1/126).
\end{align}

\paragraph{Point 3}\hfill\\
Again there is a large number of nonnegative points satisfying
$\mathbf x \cdot\mathbf{\tilde q}^{(2)} < 1$.
As an illustration of our rule that lexicographic ordering serves as a
tie-breaker if more than one ${\mathbf x}$ minimizes
$\mathbf x \cdot\mathbf{\tilde q}^{(k)}$ (\cref{item:7}b)
consider the possible choice of $\mathbf x^{(3)} = (1, 0, 0, 3, 0)$.
This is not permitted because $\mathbf x^{(3)} \cdot\mathbf{\tilde q}^{(1)} = \mathbf x^{(2)} \cdot\mathbf{\tilde q}^{(1)}$ and $\mathbf x^{(3)} > \mathbf x^{(2)}$,
thereby violating the requirement that $\mathbf x^{(2)}$ is the lexicographically largest point that minimizes $\mathbf x \cdot\mathbf{\tilde q}^{(1)}$.
An allowed choice is
\begin{align}
  \mathbf x^{(3)} = (0, 0, 0, 43, 0),
\end{align}
so that the polytope in $\mathbf q$-space becomes
\begin{align}
\begin{pmatrix}
1/3 & 55/387 & 1/774 & 1/43 & 0 \\
1/3 & 1/7 & 0 & 1/43 & 1/1806
\end{pmatrix}.\label{qlineseg}
\end{align}
As explained in \cref{sec:impl}, this is one of the pairs of 5-tuples that are gathered and processed later for performance reasons.
For this example we just continue the algorithm and find the weight system
\begin{align}
  \mathbf{\tilde q}^{(3)} = (1/3, 386/2709, 1/1548, 1/43, 1/3612).
\end{align}

\paragraph{Point 4}\hfill\\
For the final step we choose the point
\begin{align}
  \mathbf x^{(4)} = (0, 7, 0, 0, 1),
\end{align}
leading to the weight system
\begin{align}
  \mathbf{\tilde q}^{(4)} = (1/3, 571/3999, 1/7998, 1/43, 2/3999).
\end{align}
We have now collected five $n=5$, $r=1/2$ weight systems.
To be relevant for our classification of polytopes in five dimensions, we have to add a weight of one half, such that we obtain weight systems with $n=6$, $r=1$:
\begin{align}
\begin{aligned}
\mathbf{\hat q}^{(0)} &= (1/2, 1/10, 1/10, 1/10, 1/10, 1/10), \\
\mathbf{\hat q}^{(1)} &= (1/2, 1/3, 1/24, 1/24, 1/24, 1/24), \\
\mathbf{\hat q}^{(2)} &= (1/2, 1/3, 25/189, 1/54, 1/126, 1/126), \\
\mathbf{\hat q}^{(3)} &= (1/2, 1/3, 386/2709, 1/1548, 1/43, 1/3612), \\
\mathbf{\hat q}^{(4)} &= (1/2, 1/3, 571/3999, 1/7998, 1/43, 2/3999).
\end{aligned}
\label{eq:example-ws}
\end{align}
Finally, the IP check and calculation of Hodge numbers and point numbers can be performed with the result found in \cref{tab:example-ws}.
\begin{table}
  \centering
  \begin{tabular}{@{}crrrrrr@{}}
    \toprule
    weight system & $h^{1,1}$ & $h^{1,2}$ & $h^{1,3}$ & $n_p$ & $n_v$ & $n_f$ \\
    \midrule
    $\mathbf{\hat q}^{(0)}$ & 1 & 0 & 976 & 1128 & 6 & 6 \\
    $\mathbf{\hat q}^{(1)}$ & 2 & 0 & 3878 & 4551 & 6 & 6 \\
    $\mathbf{\hat q}^{(2)}$ &43 & 3 & 4884 & 5709 & 10 & 8 \\
    $\mathbf{\hat q}^{(3)}$ &912 & 0 & 43544 & 51069 & 9 & 9 \\
    $\mathbf{\hat q}^{(4)}$ & \multicolumn{3}{c}{not reflexive} & 197084 & 10 & 8 \\
    \bottomrule
  \end{tabular}
  \caption{Hodge numbers, number of points $n_p$, number of vertices $n_v$, and number of faces $n_f$ of the polytopes corresponding to the weight systems~\eqref{eq:example-ws} obtained in the example.}
  \label{tab:example-ws}
\end{table}

As a side remark we mention that, while it is fairly typical that the $\mathbf{q}$-space polytopes are simplices (as they were in the present example), this is of course not necessary.

\section{Implementation}\label{imp}
\label{sec:impl}
Our starting point was the implementation of the algorithm in PALP 2.1 \cite{math/0204356,Braun:2012vh} as it was used for constructing weight systems for reflexive Gorenstein cones \cite{Skarke:2012zg}.
While the actual classification algorithm was rewritten in C++, the check for the IP property and reflexivity was delegated to PALP's existing highly optimized C routines.
For the Hodge number computation we relied, as described below, on an improved version of PALP's C code.
The programs were compiled with \textit{UndefinedBehaviorSanitizer} enabled using the flags \texttt{-fsanitize={\allowbreak}signed-{\allowbreak}integer-{\allowbreak}overflow}
and \texttt{-fsanitize-{\allowbreak}undefined-{\allowbreak}trap-{\allowbreak}on-{\allowbreak}error}.
This ensures that signed integer arithmetic overflows are detected during run time.

After we improved redundancy avoidance along the lines indicated in the last paragraphs of section \ref{alg}, some experimentation showed that it was most efficient to apply it only at the upper levels of the recursion tree since it tended to be quite time consuming if used at every node.
With redundancy avoidance turned off at the two lowest branching levels,
it was possible to run the algorithm on a single machine down to the last branching level within 7 minutes for the case $n=6,r=1$ and within 4 minutes for the case $n=5,r=1/2$.
At that level the allowed polytope (\ref{qpol}) in $\mathbf q$-space is one-dimensional, i.e.\ it is a line segment bounded by two $n$-tuples with nonnegative entries as illustrated in (\ref{qlineseg}).
These data were sorted and residual redundancies were removed, resulting in \num{46739902} pairs of 5-tuples and \num{59048418} pairs of 6-tuples which were then distributed to 6 PCs for the last level of the recursion.
After roughly 53 hours on each machine, a total of 640 core hours, we obtained approximately $5.1 \times 10^{11}$ weight systems which amounted to 5.3\,TB of data.
The weight systems were sorted and after duplicates were removed
the result consisted of \num{108340852387} candidates for $n=6$, $r=1$ 
and \num{228960353952} candidates for $n=5$, $r=1/2$.

These weight systems still needed to be checked for the IP property and 
reflexivity;
in the reflexive case we would also want to compute the corresponding Hodge numbers.
While PALP's IP-check was made very efficient for the classification of reflexive polytopes in 4d, the Hodge number computation had not been a bottleneck so far.
In the present project, however, this was different.

In order to construct a pair of polytopes as well as the corresponding set of Hodge numbers from a weight system $\mathbf q$, the following steps have to be taken.
The lattice points (\ref{Dq}) must be enumerated and the equations describing the facets of the polytope that forms their convex hull (i.e.\ the polytope $\D_\mathbf{q}$ of eq.~(\ref{ipws}), up to the coordinate shift which we ignore here) must be computed.
PALP is good at these tasks and there was no need for improvement.
If $\D_\mathbf{q}$ is reflexive then all of its equations will correspond to integer points in the dual lattice, thereby providing the vertices of $\D_\mathbf{q}^*$.
In order to evaluate formula (\ref{baho}) for the Hodge numbers $h^{1,i}$ we also require information on the faces of $\D_\mathbf{q}$ and $\D_\mathbf{q}^*$ and on the numbers $l(\th)$ of lattice points and $l_\mathrm{int}(\th)$ of interior lattice points on a face $\th$.
PALP has efficient routines for analysing the face structure by using bit patterns \cite{math/0204356,Braun:2012vh}, which also perform well in the present context.
Finally, the point counting works as follows.
PALP creates a complete list of lattice points of $\D_\mathbf{q}^*$ by first identifying a parallelepiped $P$ that contains all the vertices of $\D_\mathbf{q}^*$ ($P$ is bounded by $n$ of the hyperplanes bounding $\D_\mathbf{q}^*$ as well as hyperplanes parallel to these), checking for every lattice point of $P$ whether it belongs to $\D_\mathbf{q}^*$, and adding such a point to the list if it does;
examining all lattice points of $P$ corresponds to a nested set of loops in the program.
Then PALP goes over the complete list of points and the complete list of faces and raises $l_\mathrm{int}$ of the face if appropriate.
Here we achieved a considerable upgrade of 
efficiency.
We improved the conditions upon which the program exits from a particular loop.
Furthermore, instead of creating the full set of lattice points of $\D_\mathbf{q}^*$ we used the following trick.
The innermost loop level corresponds to a sequence of lattice points along a line in $\D_\mathbf{q}^*$.
Having worked out to which faces the first and last point in the line are interior, all other points must be interior to the affine span of these faces, and we can immediately raise the corresponding numbers $l_\mathrm{int}$ with the appropriate multiplicities, without having to create and analyse the full list.

Despite these improvements it would have taken a long time to process all candidate weight systems on our local computers.
We therefore used the facilities of the Vienna Scientific Cluster (VSC-3).
The IP check and Hodge number calculation for the \num{337301206339} weight system candidates was distributed among 119 nodes and completed in \num{57321} core hours on machines with Intel Xeon E5-2650 v2 processors clocked at 2.6\,GHz.

Weight systems that determine reflexive polytopes were sorted according to their Hodge numbers, the ones leading to non-reflexive polytopes according to vertex count, facet count, and point count.
All of them were stored in a PostgreSQL database.

Then we compared our results with the two largest existing lists of weight systems. 
One of them is the complete list of IP weight systems with $\sum_{i=1}^6 q_i \le 300$ (in the normalization in which the $q_i$ are integer) which was generated by considering suitable partitions of the numbers up to 300 (to be found at the website \cite{cydata}; it represents a straightforward generalization of the list up to $\sum_{i=1}^6 q_i \le 150$ presented in \cite{Kreuzer:1997zg}).
The other one was the complete list of \num{1100055} hypersurfaces in weighted
projective spaces \cite{hep-th/9812195}.
We confirmed that every single weight system occurring in either of these lists could be found in the database.
Since the construction of our database was completely independent both conceptually and computationally from the way these lists were generated, it seems very unlikely that there is an error in our algorithm or programming.
Together with the fact that we excluded the possibility of numerical errors from overflows, which might have led to misinterpreting viable weight systems as non-IP, this gives us quite an amount of confidence in the reliability of our results.

\section{Results and discussion}\label{res}


There are \num{322383760930} weight systems with six weights that determine five-dimensional polytopes with the IP property.
\num{185269499015} of these polyhedra are reflexive, \num{137114261915} non-reflexive.
The PostgreSQL database which contains all of these data is searchable via a web front-end at: \\

\centerline{\url{http://rgc.itp.tuwien.ac.at/fourfolds}}\hfill

The reflexive polytopes give rise to \num{532600483} distinct sets of Hodge numbers.
Thus a Hodge number triple in our list occurs on average for roughly 350 weight systems.
This number is, of course, just the mean of a strongly skewed distribution.
The Hodge data sets with the highest numbers of occurrences are shown in
Table \ref{topten}.

\begin{table}
\centering
\begin{tabular}{@{}rrrrrrr@{}}
\toprule
&    $h^{1,1}$  &  $h^{1,2}$  &  $h^{1,3}$  &   $h^{2,2}$   &  $\chi$   & weight system count \\
\midrule
1 &   \num{25827} &   0 &  13 & \num{103404} & \num{155088} & \num{660386443} \\
2 &   \num{28348} &   0 &  12 & \num{113484} & \num{170208} & \num{650642665} \\
3 &   \num{23426} &   0 &  14 &  \num{93804} & \num{140688} & \num{388024998} \\
4 &   \num{22386} &   0 &  15 &  \num{89648} & \num{134454} & \num{323589412} \\
5 &   \num{30989} &   0 &  11 & \num{124044} & \num{186048} & \num{289288747} \\
6 &   \num{24738} &   0 &  14 &  \num{99052} & \num{148560} & \num{239597804} \\
7 &   \num{23946} &   0 &  14 &  \num{95884} & \num{143808} & \num{230489503} \\
8 &   \num{25746} &   0 &  14 & \num{103084} & \num{154608} & \num{211084163} \\
9 &   \num{27548} &   0 &  12 & \num{110284} & \num{165408} & \num{193560096} \\
10 &  \num{20154} &   0 &  16 &  \num{80724} & \num{121068} & \num{190167835} \\
\bottomrule
\end{tabular}
\caption{The ten most frequent sets of Hodge numbers.}
\label{topten}
\end{table}
One should note that distinct weight systems may well lead to the same polytope
(we have not checked how often this occurs).
In particular it seems that
polytopes with a small number of lattice points are generated many times,
which accounts for the fact that $h^{1,1}\gg h^{1,3}$ for all the entries of
Table \ref{topten}.
The Hodge numbers $h^{i,j}$ and the Euler characteristic $\chi$ of the
reflexive polyhedra lie within the following ranges:
\begin{itemize}
\item 1 $\le h^{1,1} \le$ \num{303148} (with \num{190201} distinct values),
\item 0 $\le h^{1,2} \le$ \num{2010} (with \num{1689} distinct values),
\item 1 $\le h^{1,3} \le$ \num{303148} (with \num{145848} distinct values),
\item 82 $\le h^{2,2} \le$ \num{1213644} (with \num{361426} distinct values),
\item -252 $\le \chi \le$ \num{1820448} (with \num{188804} distinct values).
\end{itemize}

The appendix to this paper contains a number of figures with which we try to
visualize our data.
Because of formula (\ref{h22formula}) and the standard dependence of $\chi$ on
the Hodge numbers, the space of quintuples
$(h^{1,1}, h^{1,2}, h^{1,3}, h^{2,2}, \chi)$ is really a three-dimensional data set.
Due to the size of this set we found no way of adequately visualizing it in
its full dimensionality.
Instead we have mainly relied on the two-dimensional
plot of $(h^{1,1},h^{1,3})$, i.e.\ numbers of Kähler and complex structure
moduli, which is the straightforward generalization of the usual Hodge
number plot for threefolds.
It is also the most natural choice in the sense that the missing direction is the one along which our data set is thinnest (as one sees from the list above, the ranges for $h^{1,1}$ and $h^{1,3}$ are larger by a factor of $\sim 150$ than that for $h^{1,2}$).

\begin{figure}[H]
  \centering
  \includegraphics{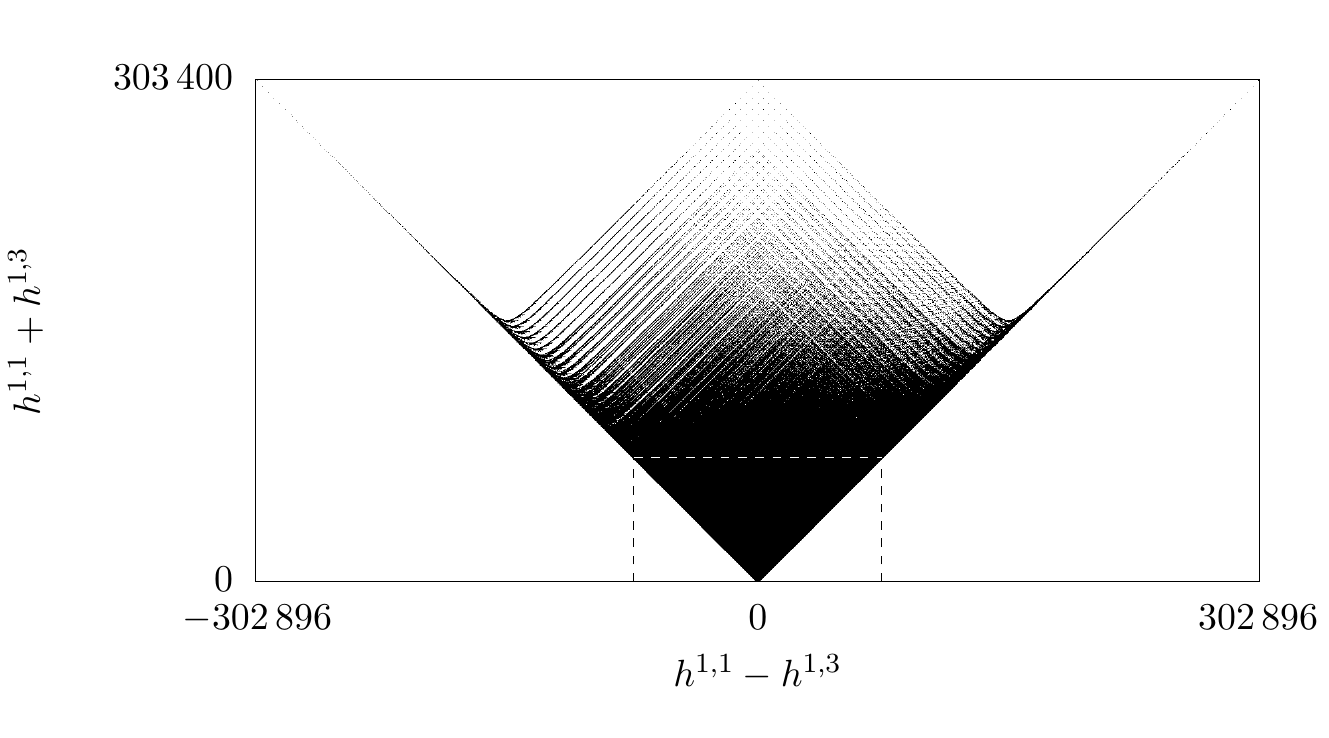}
  \caption{All values of $(h^{1,1},h^{1,3})$; the rectangle bounded by dashed lines indicates the range of Fig.~\ref{hodgeplot2}.}
  \label{hodgeplotall}
\end{figure}

Fig.~\ref{hodgeplotall} presents the shape of the whole dataset.
Similarly to the corresponding set for threefolds, it is dense (in the sense that every possible pair occurs) in a large region with moderate values of $h^{1,1}$ and $h^{1,3}$ and shows a characteristic symmetric shape with three peaks and a grid structure related to fibrations whose fibres correspond to self-dual polyhedra of one dimension less \cite{ARXIV:1207.4792}.
Apparently the set of tips in any dimension can be described in the following manner.
Consider the sequence of integers
\beq (a_i) = (2,3,7,43,1807,\ldots ) \eeq
generated by the rule
\beq a_1=2, \quad a_{n+1} = 1 + \prod_{i=1}^n a_i = a_n(a_n-1) + 1.  \eeq
It is easy to show by induction that
\beq \sum_{i=1}^n {1\over a_i} = 1 - {1\over \prod_{i=1}^n a_i},  \eeq
which implies that the $n$-tuples
\bea {\mathbf q}_\mathrm{ct}^{(n)} &=&
\({1\over a_1},\ldots,{1\over a_{n-1}},{1\over \prod_{i=1}^{n-1}a_i}\),\\
{\mathbf q}_\mathrm{lt}^{(n)} &=&
\({1\over a_1},\ldots,{1\over a_{n-2}},{1\over 2 \prod_{i=1}^{n-2}a_i},{1\over 2 \prod_{i=1}^{n-2}a_i}\)
\eea
form weight systems with $\sum q_i = 1$.
Since each weight is the inverse of an integer (i.e.\ they are ``Fermat weights'') both of these weight systems have the IP property.
Comparison with our data shows that
\beq {\mathbf q}_\mathrm{ct}^{(6)} = \( \frac{1}{2}, \frac{1}{3}, \frac{1}{7},
\frac{1}{43}, \frac{1}{1807}, \frac{1}{3263442} \) ~\mathrm{ and }~
{\mathbf q}_\mathrm{lt}^{(6)} = \( \frac{1}{2}, \frac{1}{3}, \frac{1}{7}, \frac{1}{43}, \frac{1}{3612}, \frac{1}{3612} \) \nn\eeq
correspond to the central upper tip and to the left tip in Fig.~\ref{hodgeplotall}, respectively.
The analogous statements for Calabi--Yau threefolds ($n=5$) are also easily checked.
The right upper tip corresponds of course to the polytope that is dual to the one determined by ${\mathbf q}_\mathrm{lt}$.
If we represent $\D_{{\mathbf q}_\mathrm{lt}^{(n)}}$ and
$\D_{{\mathbf q}_\mathrm{ct}^{(n)}}$ as in (\ref{ipws}), so that the interior lattice
point is
$(1,\ldots,1)$, it is easy to see that the intersection of either of these
polytopes with the hyperplane $x_{n-1}=x_n$ is isomorphic to
$\D_{{\mathbf q}_\mathrm{ct}^{(n-1)}}$; likewise $\D_{{\mathbf q}_\mathrm{lt}^{(n)}}^*$, which
is isomorphic to $\D_{{\mathbf q}_\mathrm{lt}^{(n)}}$ up to a change of lattice, has
$\D_{{\mathbf q}_\mathrm{ct}^{(n-1)}}$ as a subpolytope.
These inclusions of reflexive polyhedra give rise to fibration structures where the
fibre is the self-mirror Calabi--Yau manifold of dimension one less that
corresponds to $\D_{{\mathbf q}_\mathrm{ct}^{(n-1)}}$; see Ref.~\cite{ARXIV:1207.4792}
for details of this construction and how it can be used to explain the structure
of the uppermost part of the Hodge number plot.
A further fibration structure comes from the fact that
$\D_{{\mathbf q}_\mathrm{lt}^{(n)}}$ has a subpolytope isomorphic to
$\D_{{\mathbf q}_\mathrm{lt}^{(n-1)}}$ in the hyperplane $2x_{n-2}=x_{n-1}+x_n$.
In the case of $n=4$ the two inclusions
$\D_{(1/2,\,1/3,\,1/12,\,1/12)} \supset \D_{(1/2,\,1/3,\,1/6)}$ and
$\D_{(1/2,\,1/3,\,1/12,\,1/12)} \supset \D_{(1/2,\,1/4,\,1/4)}$ correspond to distinct
elliptic fibration structures of a K3 manifold that are related to
$E_8\times E_8$ \cite{Candelas:1996su}
and $SO(32)$ \cite{Candelas:1997pq}, respectively;
via further nested inclusions these elliptic fibration structures occur in
higher dimensions as well.

\del
Each of the polytopes $\D_{{\mathbf q}_\mathrm{lt}^{(n)}}$,
$\D_{{\mathbf q}_\mathrm{lt}^{(n)}}^*$ and
$\D_{{\mathbf q}_\mathrm{ct}^{(n)}}=\D_{{\mathbf q}_\mathrm{ct}^{(n)}}^*$
has a slice through the interior point that is isomorphic to
$\D_{{\mathbf q}_\mathrm{ct}^{(n-1)}}$.
\enddel

\begin{figure}[H]
  \centering
  \includegraphics{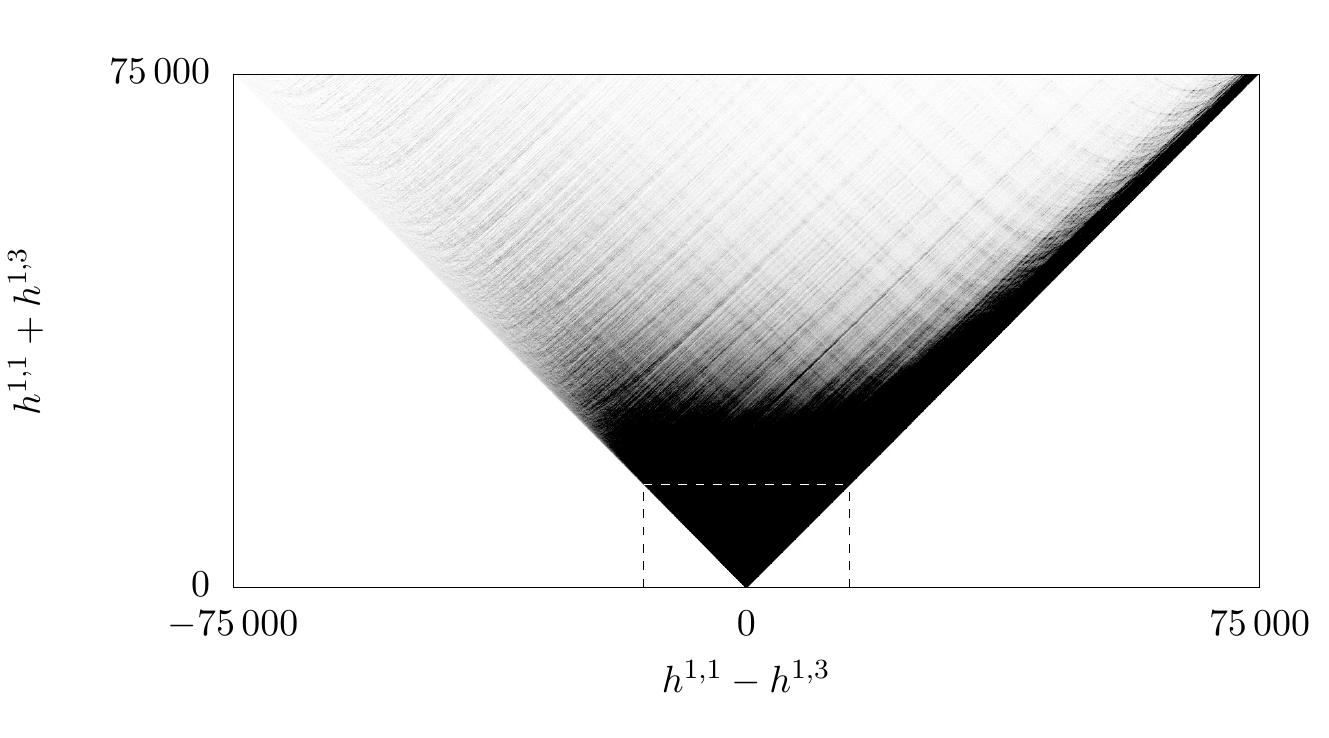}
  \caption{$(h^{1,1},h^{1,3})$ with $h^{1,1}+h^{1,3}\le \num{75000}$: greyscale indicates frequency; the rectangle bounded by dashed lines indicates the range of Fig.~\ref{hodgeplot3}. }
  \label{hodgeplot2}
\end{figure}

\begin{figure}[H]
  \centering
  \includegraphics{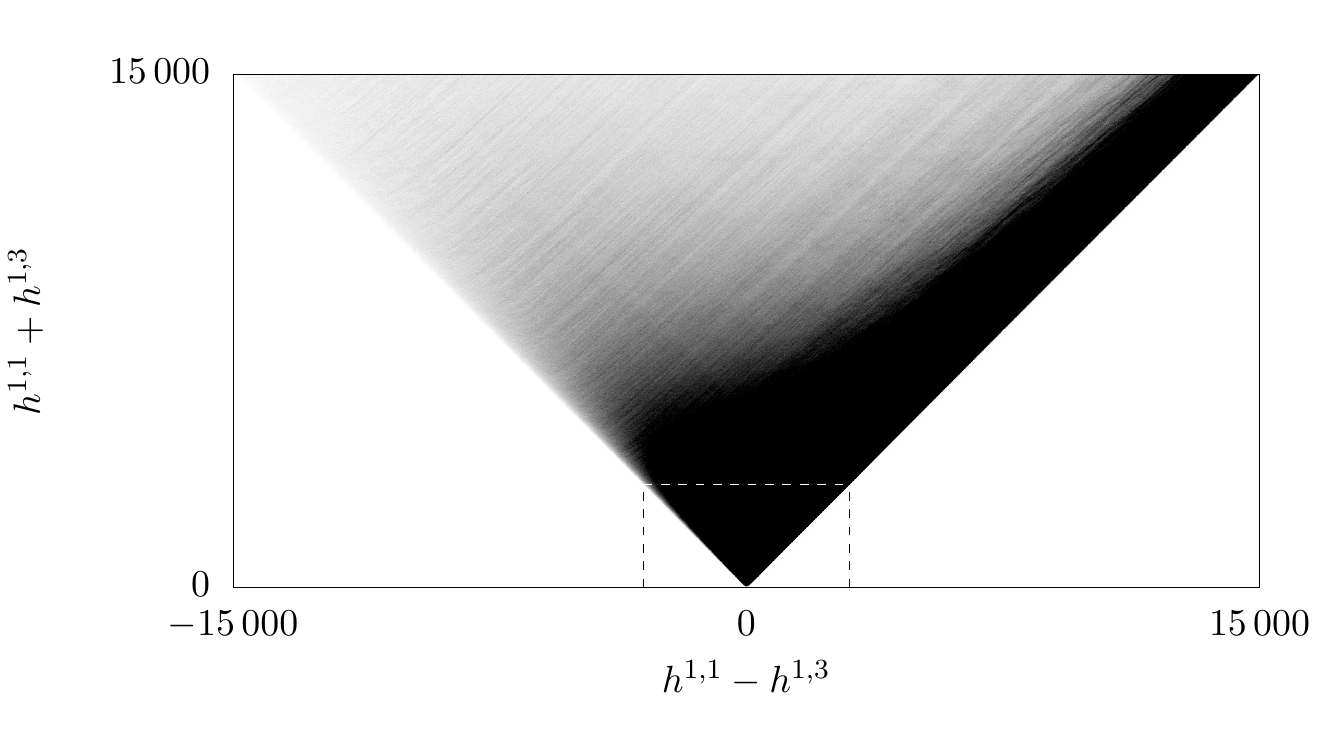}
  \caption{$(h^{1,1},h^{1,3})$ with $h^{1,1}+h^{1,3}\le \num{15000}$: greyscale indicates frequency; the rectangle bounded by dashed lines indicates the range of Fig.~\ref{hodgeplot4}.}
  \label{hodgeplot3}
\end{figure}

\begin{figure}[H]
  \centering
  \includegraphics{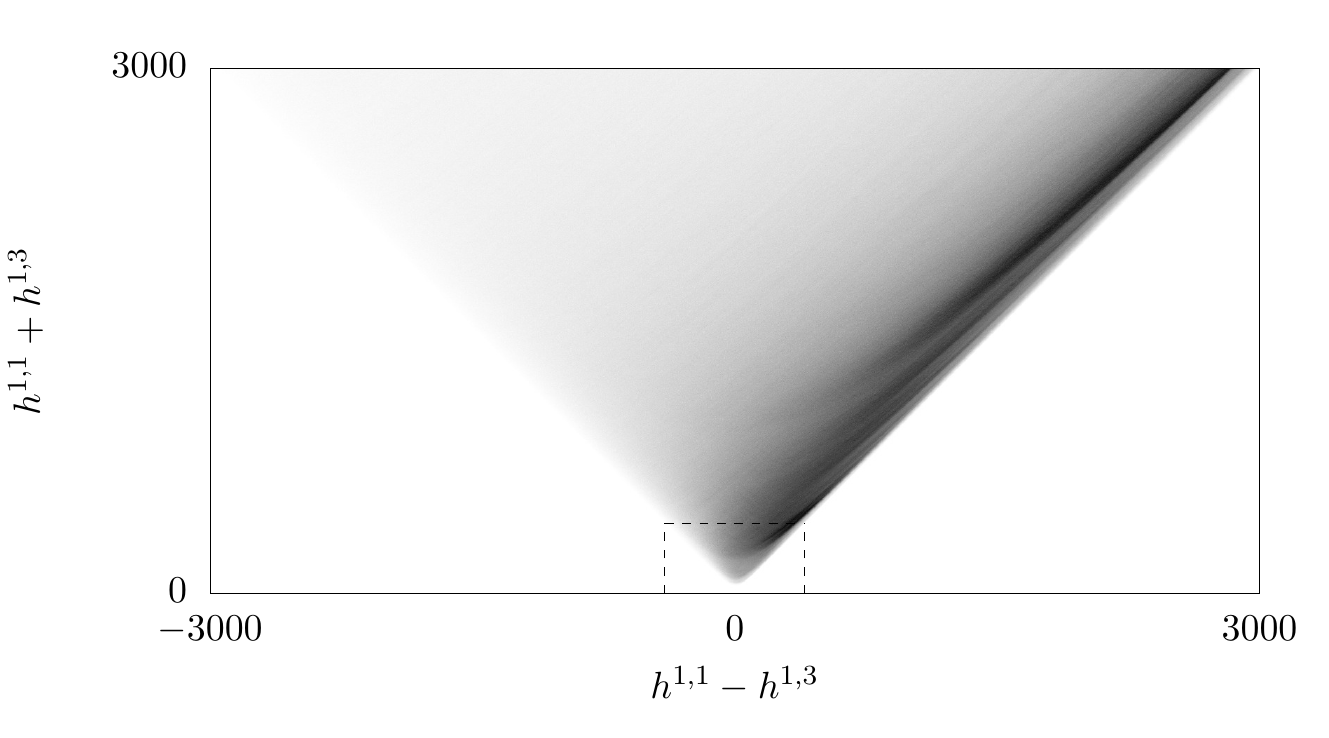}
  \caption{$(h^{1,1},h^{1,3})$ with $h^{1,1}+h^{1,3}\le \num{3000}$: greyscale indicates frequency; the rectangle bounded by dashed lines indicates the range of Fig.~\ref{hodgeplot5}. }
  \label{hodgeplot4}
\end{figure}

\begin{figure}[H]
  \centering
  \includegraphics{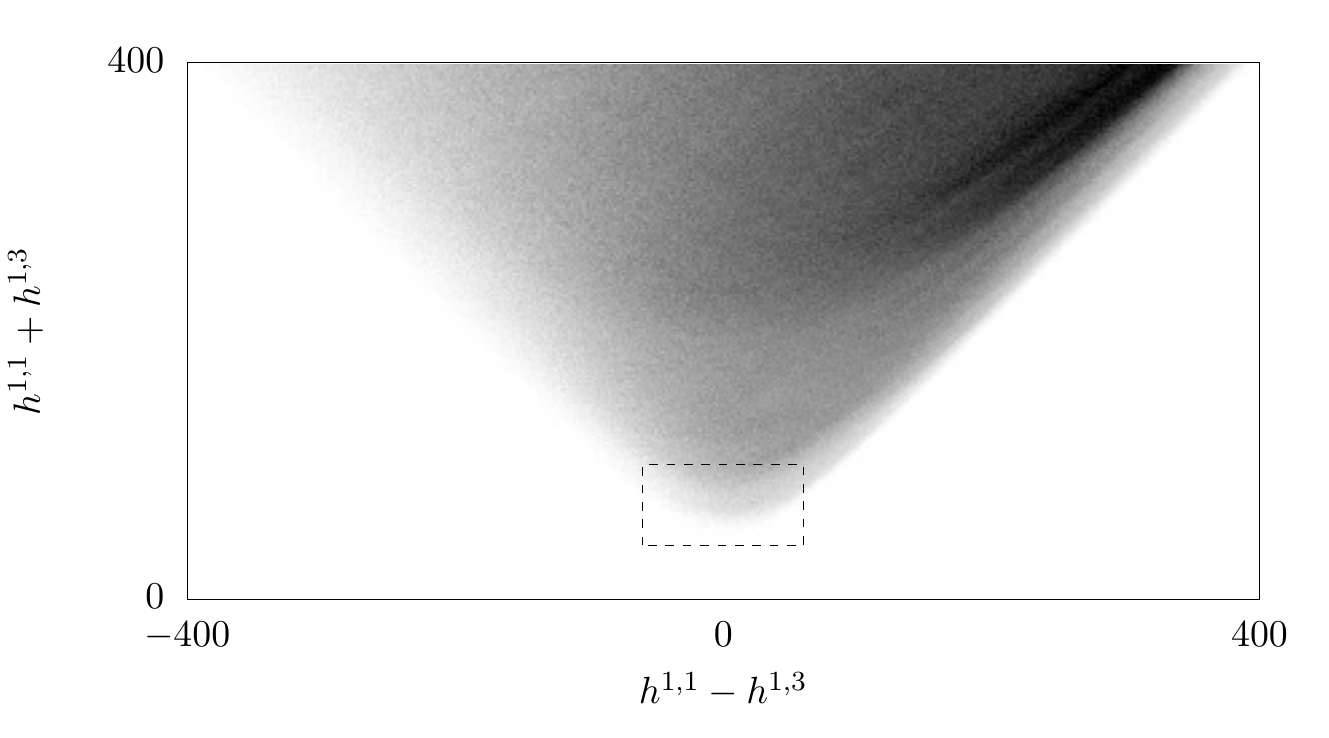}
  \caption{$(h^{1,1},h^{1,3})$ with $h^{1,1}+h^{1,3}\le 400$: greyscale indicates frequency; the rectangle bounded by dashed lines indicates the range of Fig.~\ref{hodgeplot6}. }
  \label{hodgeplot5}
\end{figure}

\begin{figure}[H]
  \centering
  \includegraphics{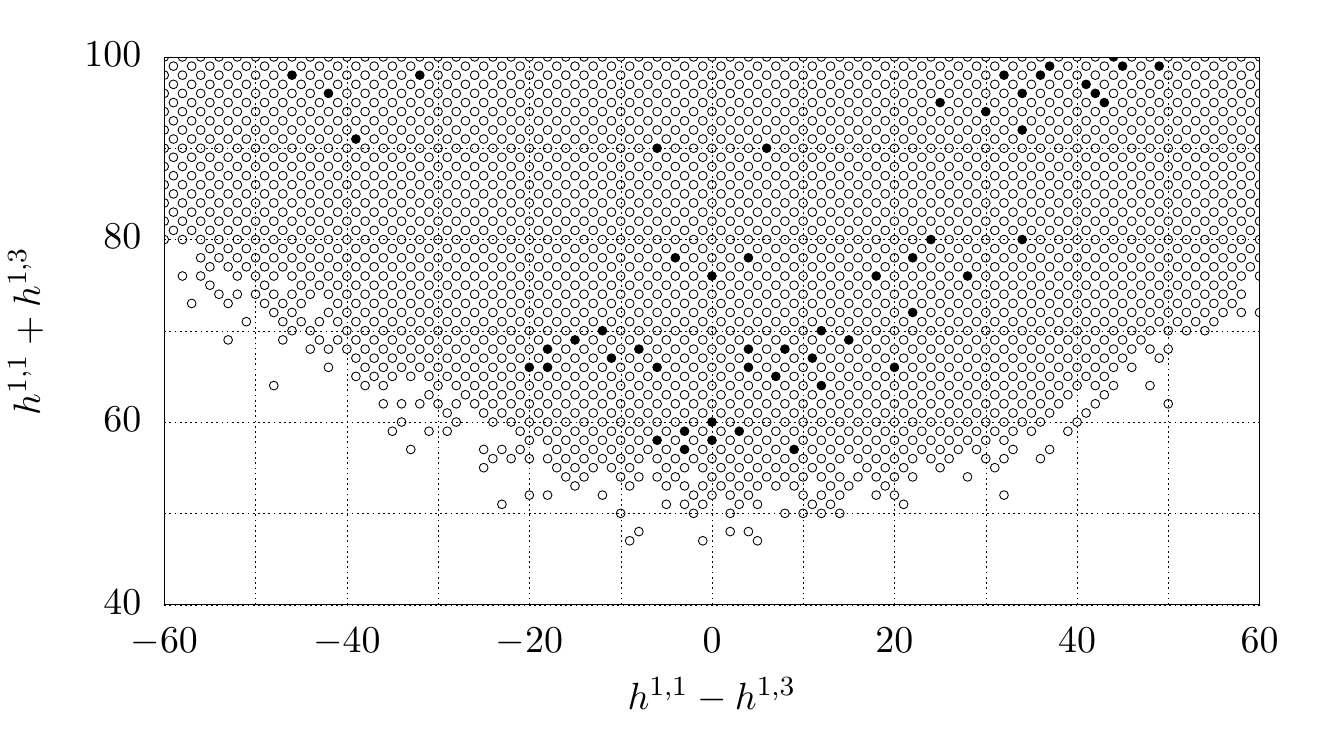}
  \caption{Small values of $(h^{1,1},h^{1,3})$: 
    solid circles indicate pairs admitting negative Euler characteristic.}
  \label{hodgeplot6}
\end{figure}

Among \cref{hodgeplot2,hodgeplot3,hodgeplot4,hodgeplot5,hodgeplot6}, each corresponds to the small
subregion of the previous plot that is indicated by the rectangle bounded by dashed lines.
With the exception of Fig.~\ref{hodgeplot6} it is impossible to display single data points as such.
Instead one should think of each pixel in Fig.~\ref{hodgeplotall} as representing information on whether or not it contains a data point.
In \cref{hodgeplot2,hodgeplot3,hodgeplot4,hodgeplot5} every pixel is given a particular shade of grey depending on the number of Hodge data sets giving rise to data points lying there; the greyscales are obviously different for different figures.
Only in Fig.~\ref{hodgeplot6} single data points are visible.
Here every pair $(h^{1,1},h^{1,3})$ that is realized by at least one weight system is indicated by a circle.
This circle is filled in those rare  cases in which a Hodge triple with negative Euler number exists (we discuss the scarcity of such points below).

\begin{figure}[H]
  \centering
  \includegraphics{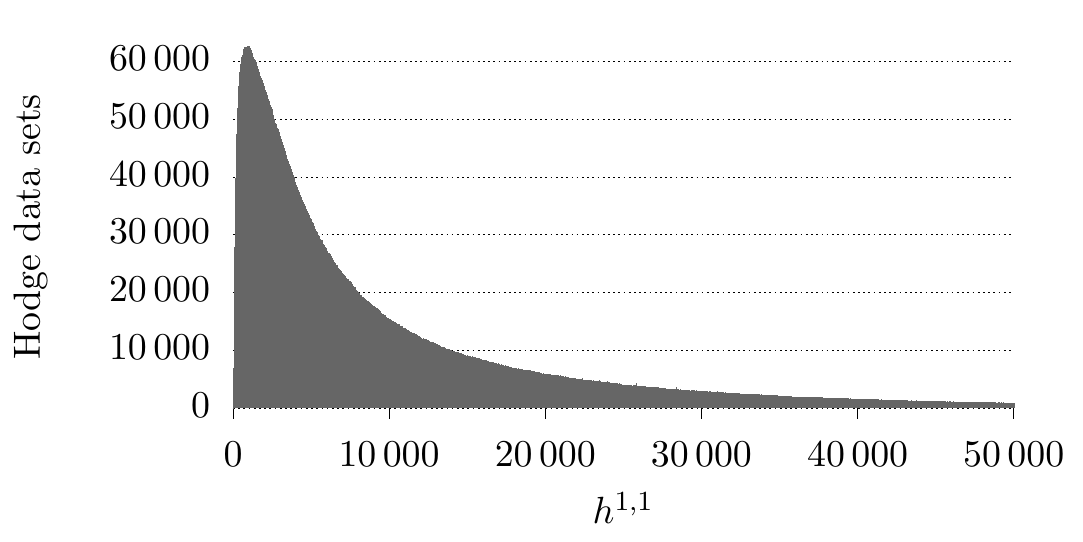}
  \caption{Numbers of Hodge data sets with a given value of $h^{1,1}$}
  \label{fig:h11-hodge-1}
\end{figure}

\begin{figure}[H]
  \centering
  \includegraphics{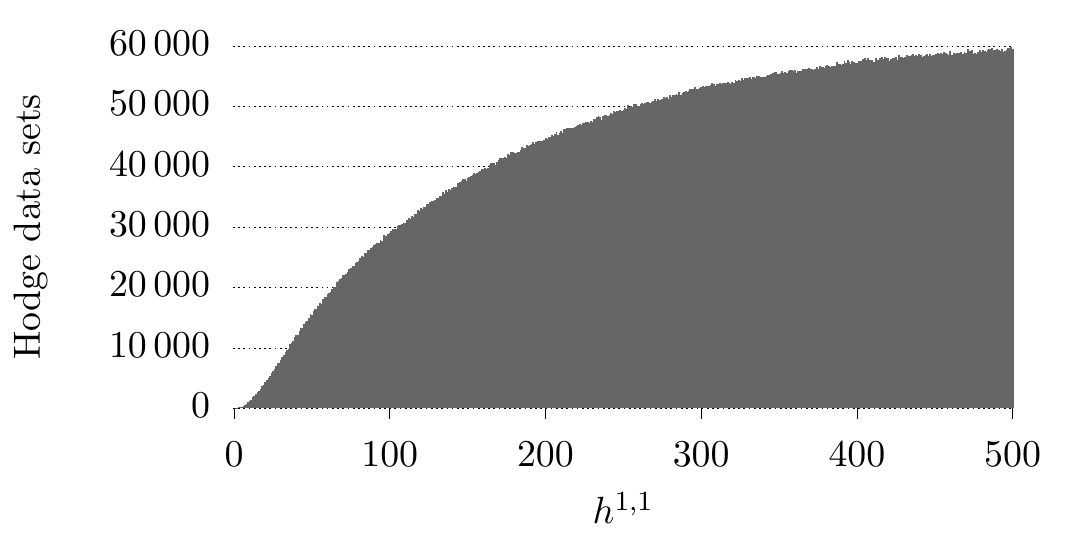}
  \caption{Numbers of Hodge data sets with a given value of $h^{1,1}$}
  \label{fig:h11-hodge-2}
\end{figure}

\begin{figure}[H]
  \centering
  \includegraphics{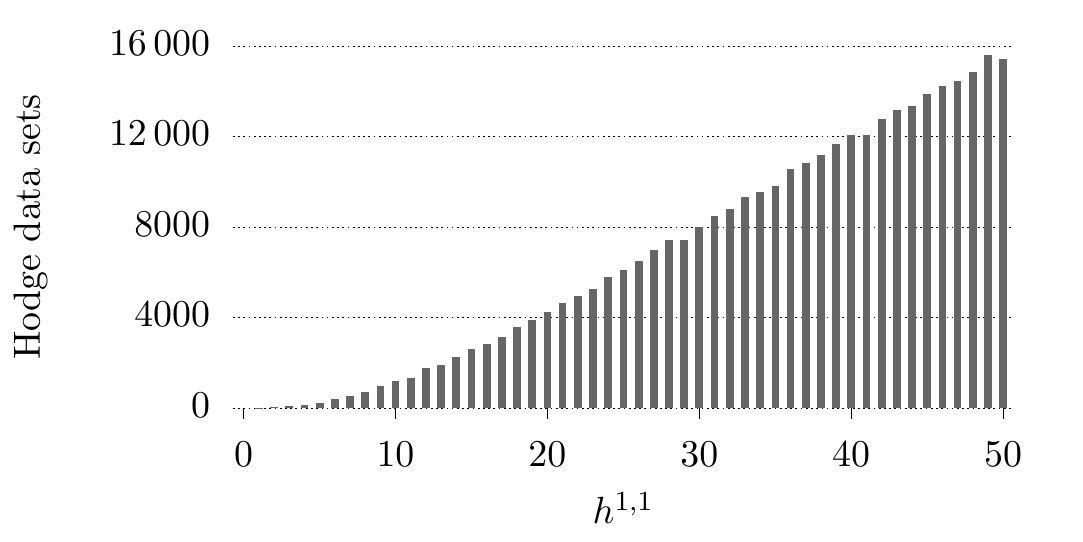}
  \caption{Numbers of Hodge data sets with a given value of $h^{1,1}$}
  \label{fig:h11-hodge-3}
\end{figure}

\begin{figure}[H]
  \centering
  \includegraphics{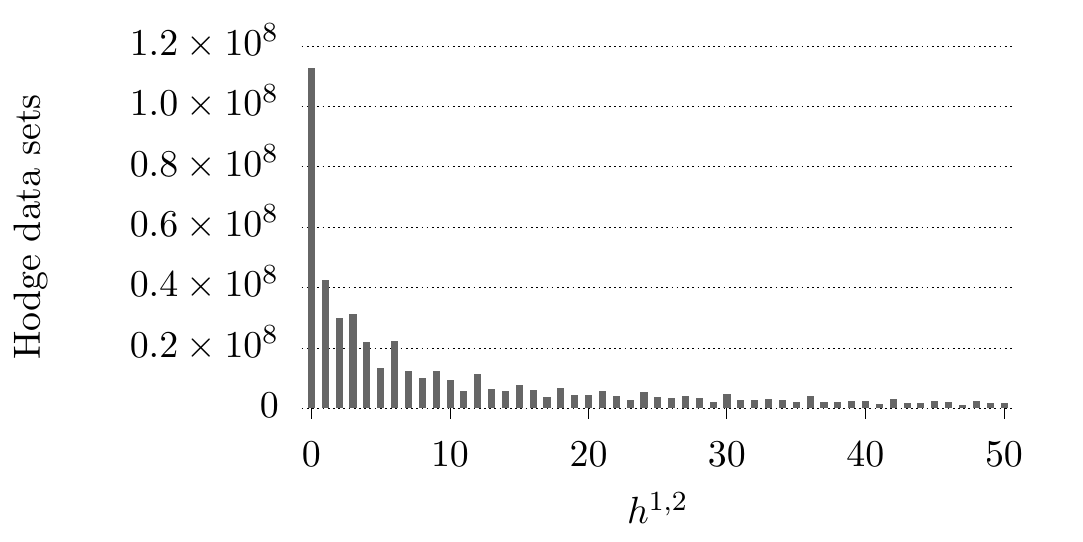}
  \caption{Numbers of Hodge data sets with a given value of $h^{1,2}$}
  \label{fig:h12-hodge-1}
\end{figure}

\begin{figure}[H]
  \centering
  \includegraphics{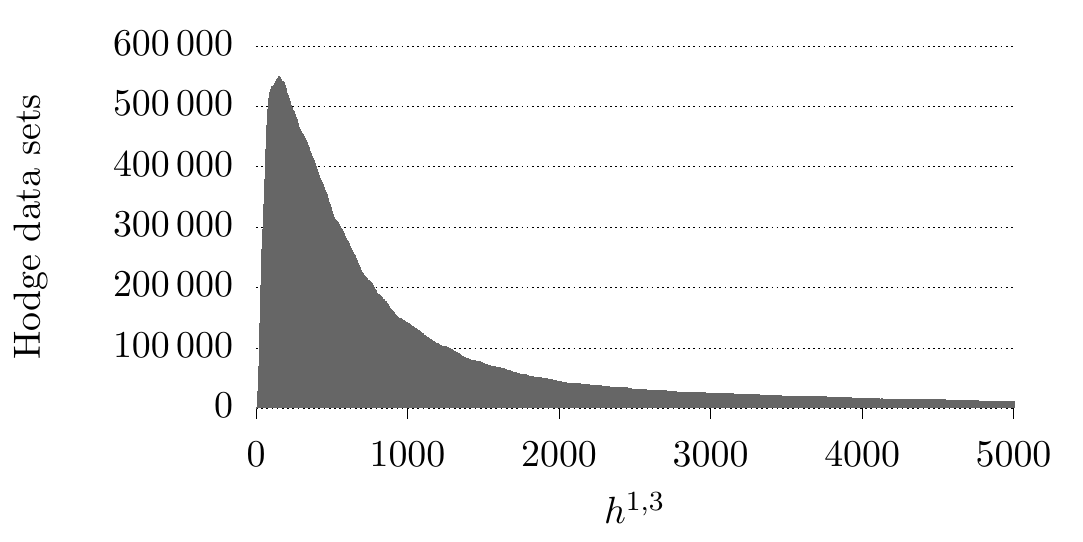}
  \caption{Numbers of Hodge data sets with a given value of $h^{1,3}$}
  \label{fig:h13-hodge-1}
\end{figure}

\begin{figure}[H]
  \centering
  \includegraphics{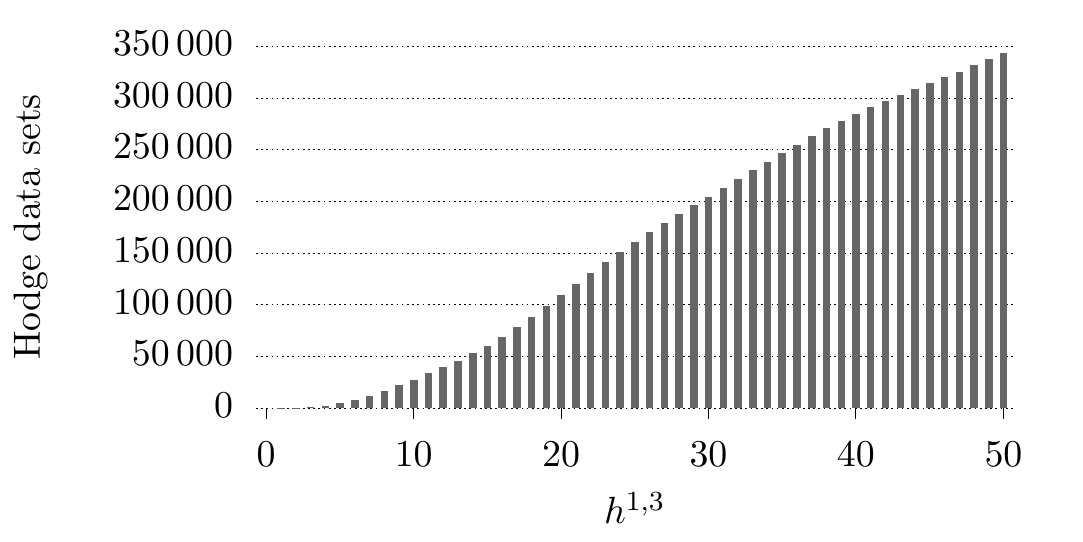}
  \caption{Numbers of Hodge data sets with a given value of $h^{1,3}$}
  \label{fig:h13-hodge-3}
\end{figure}

\begin{figure}[H]
  \centering
  \includegraphics{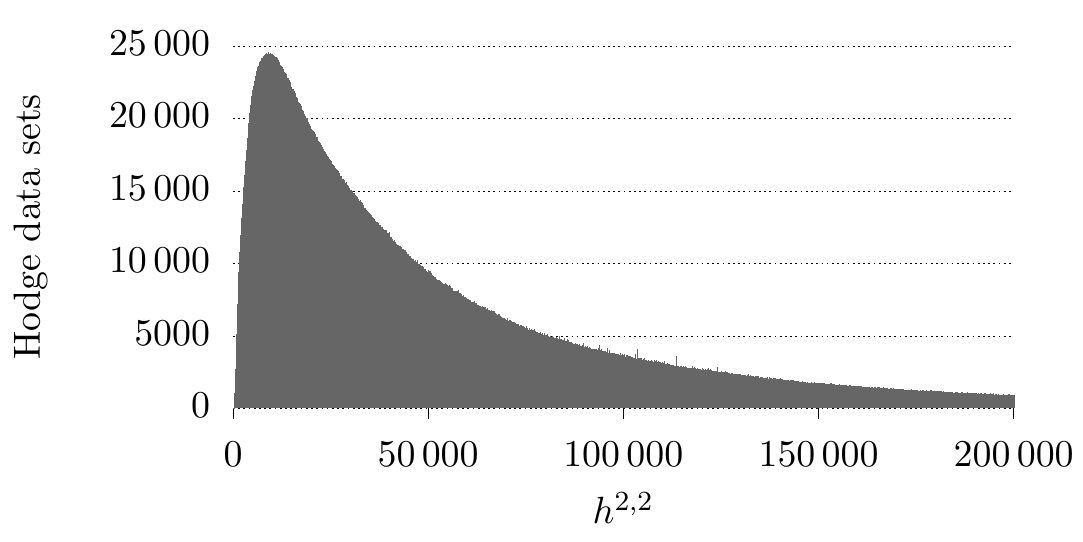}
  \caption{Numbers of Hodge data sets with a given value of $h^{2,2}$}
  \label{fig:h22-hodge-1}
\end{figure}

\begin{figure}[H]
  \centering
  \includegraphics{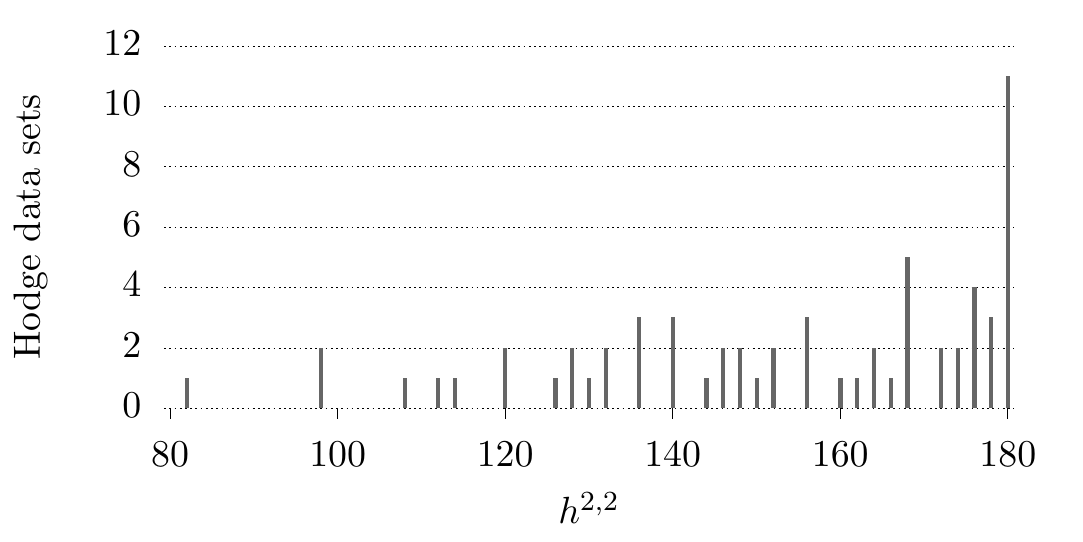}
  \caption{Numbers of Hodge data sets with a given value of $h^{2,2}$}
  \label{fig:h22-hodge-3}
\end{figure}

\begin{figure}[H]
  \centering
  \includegraphics{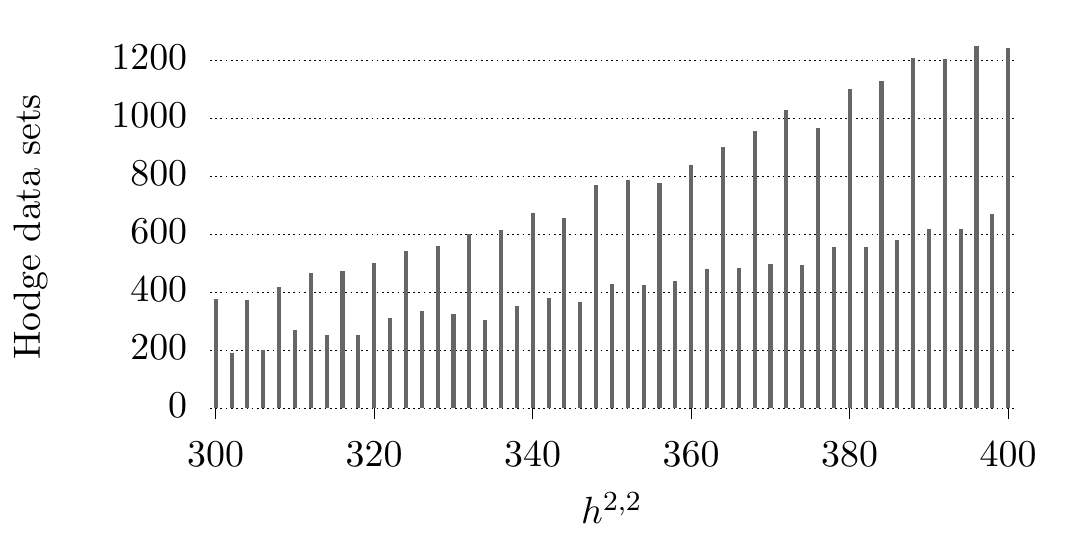}
  \caption{Numbers of Hodge data sets with a given value of $h^{2,2}$}
  \label{fig:h22-hodge-2}
\end{figure}

\begin{figure}[H]
  \centering
  \includegraphics{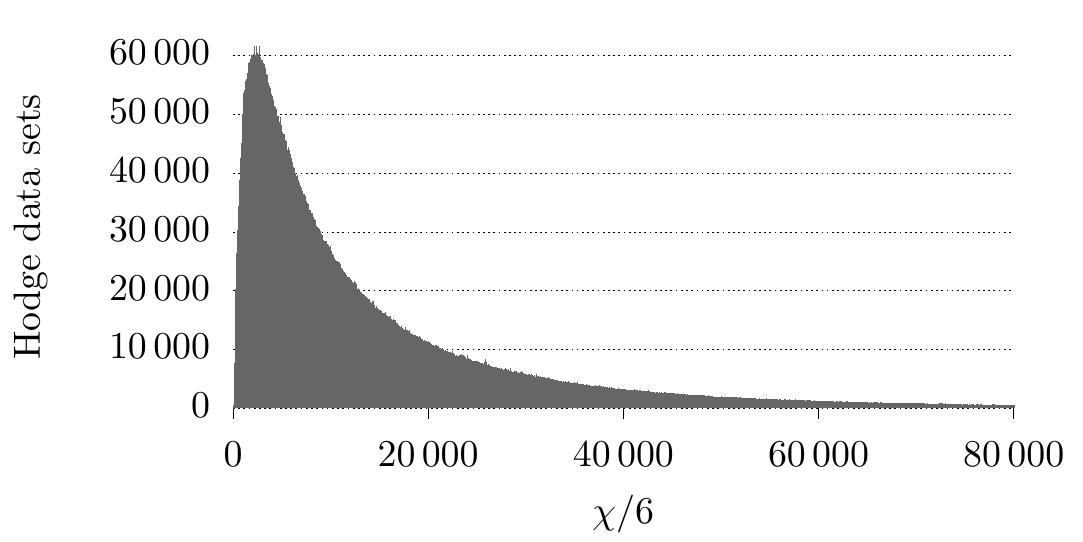}
  \caption{Numbers of Hodge data sets with a given value of $\chi$}
  \label{fig:chi-hodge-1}
\end{figure}

\begin{figure}[H]
  \centering
  \includegraphics{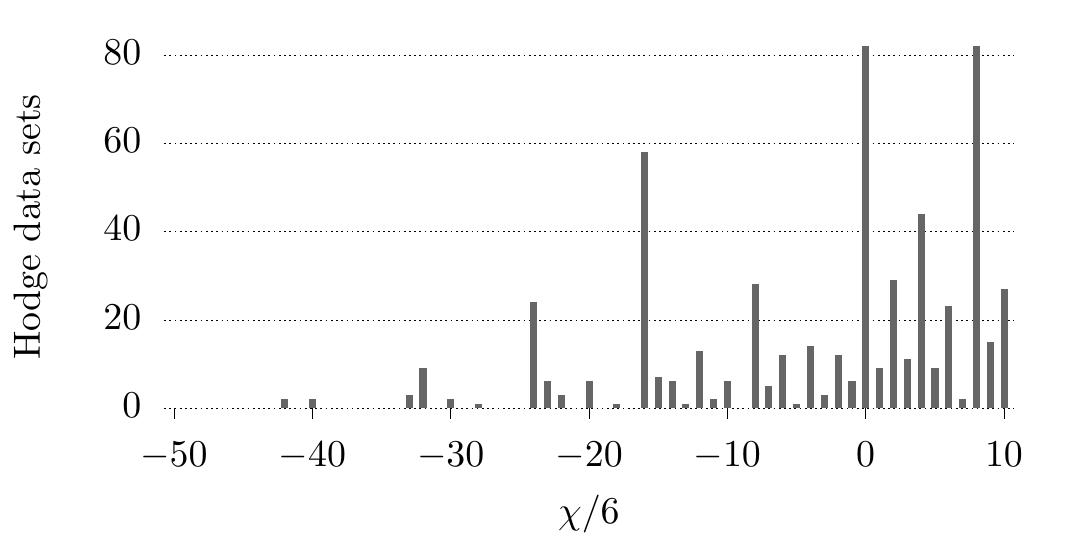}
  \caption{Numbers of Hodge data sets with a given value of $\chi$}
  \label{fig:chi-hodge-3}
\end{figure}

\begin{figure}[H]
  \centering
  \includegraphics{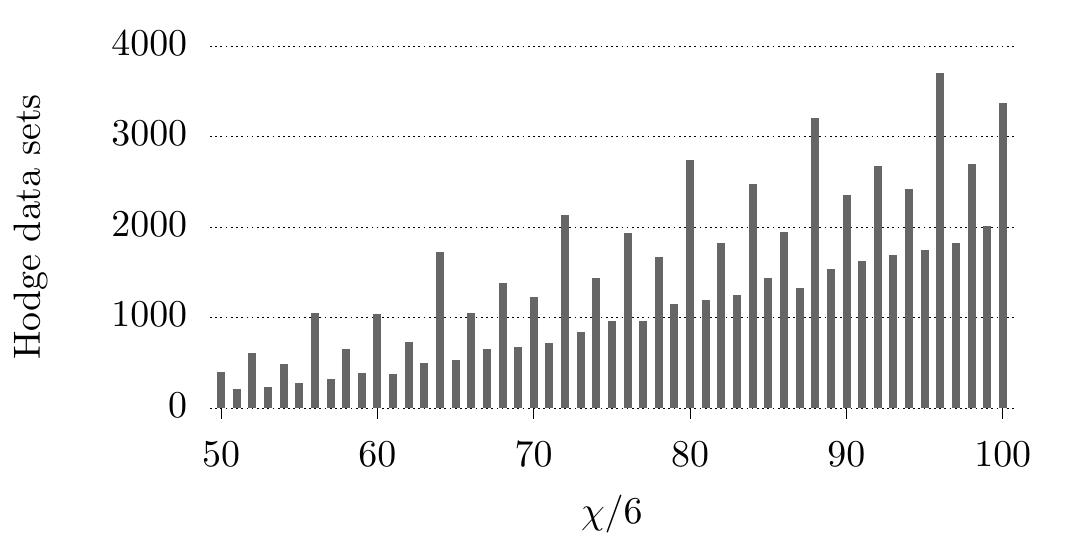}
  \caption{Numbers of Hodge data sets with a given value of $\chi$}
  \label{fig:chi-hodge-2}
\end{figure}

\begin{figure}[H]
  \centering
  \includegraphics{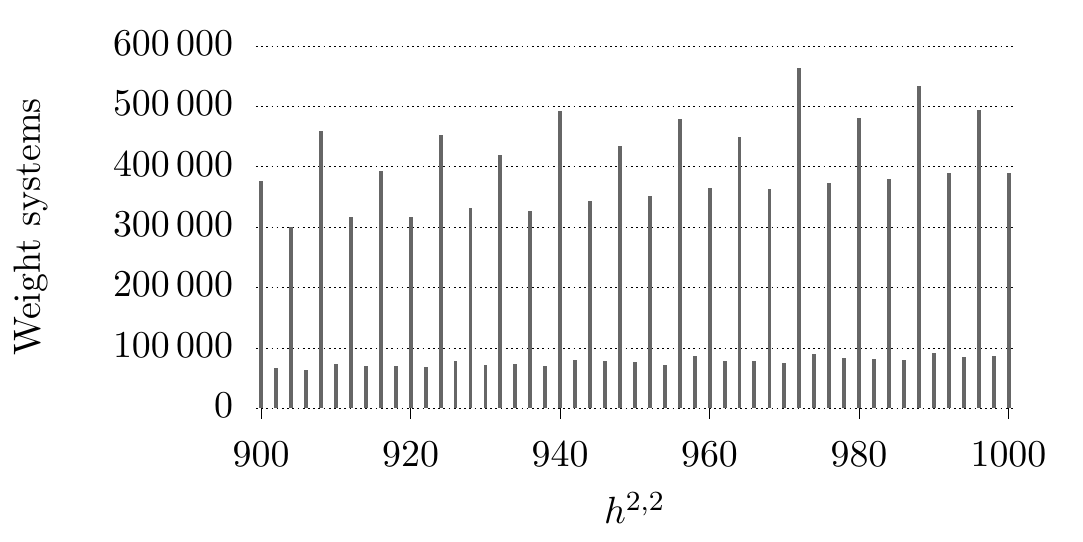}
  \caption{Numbers of weight systems leading to a given value of $h^{2,2}$}
  \label{fig:h22-ws-2}
\end{figure}

\begin{figure}[H]
  \centering
  \includegraphics{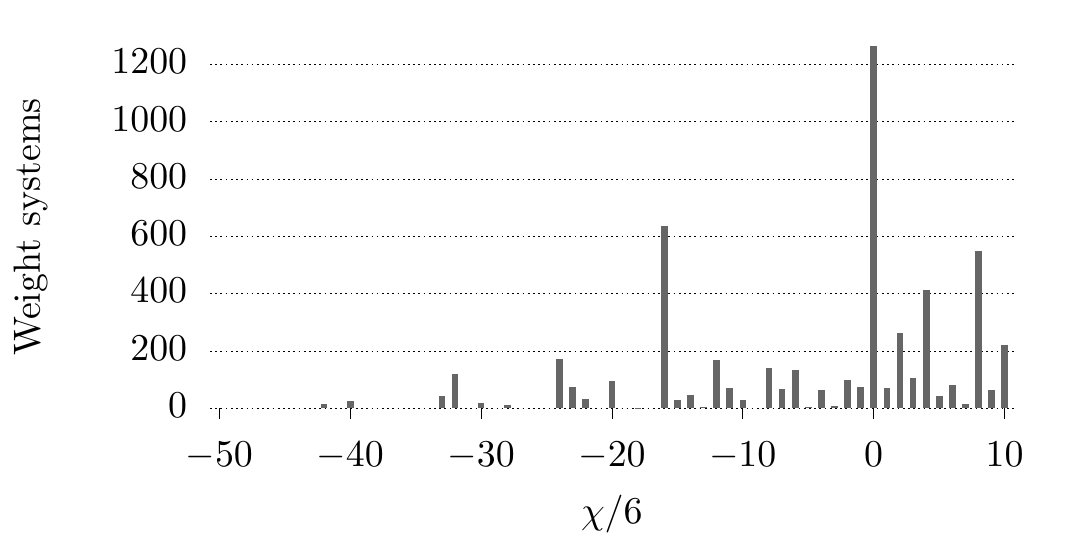}
  \caption{Numbers of weight systems leading to a given value of $\chi$}
  \label{fig:chi-ws-3}
\end{figure}

\begin{figure}[H]
  \centering
  \includegraphics{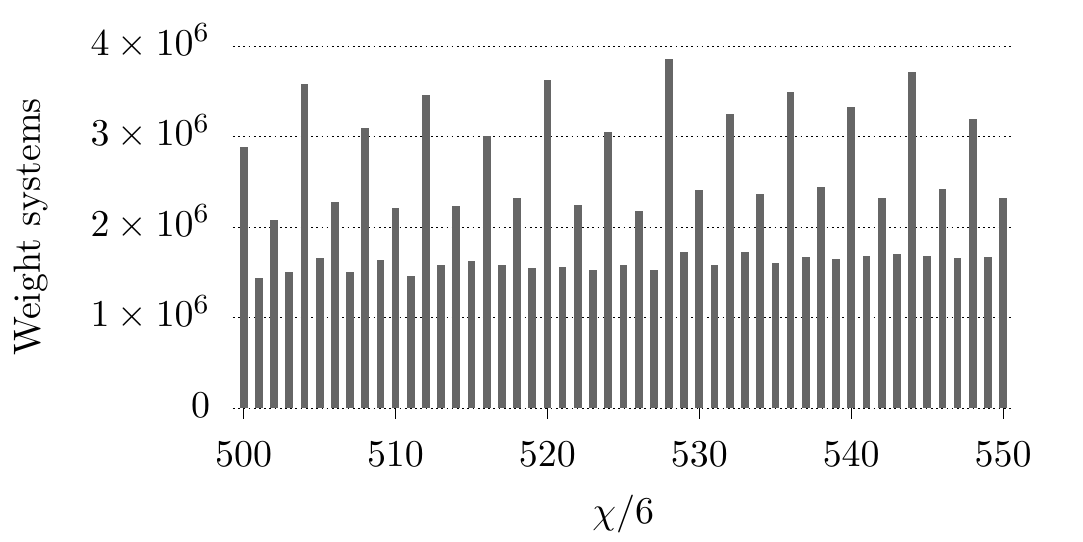}
  \caption{Numbers of weight systems leading to a given value of $\chi$}
  \label{fig:chi-ws-2}
\end{figure}

The remaining figures give information on the frequencies of occurrences of specific values.
The plots in \cref{fig:h11-hodge-1,fig:h11-hodge-2,fig:h11-hodge-3,fig:h12-hodge-1,fig:h13-hodge-1,fig:h13-hodge-3,fig:h22-hodge-1,fig:h22-hodge-2,fig:h22-hodge-3,fig:chi-hodge-1,fig:chi-hodge-2,fig:chi-hodge-3} indicate the numbers of Hodge data sets in which a particular value of one of the Hodge numbers or $\chi$ is taken, whereas the remaining plots indicate how many different weight systems give rise to the quantity in question.
Perhaps the most notable feature of these plots is the distribution of possible
values for the Euler characteristic $\chi$.
Unless one zooms into the very left end of the distribution, as in Fig.~\ref{fig:chi-hodge-3} and Fig.~\ref{fig:chi-ws-3}, the plots appear to start at $\chi = 0$.
The somewhat surprising scarcity and small values of negative Euler characteristics are consequences of 
\beq \chi = 4 + 2 h^{1,1} - 4 h^{1,2} + 2 h^{1,3} + h^{2,2}
= 6(8 + h^{1,1} - h^{1,2} + h^{1,3}) \eeql{chi}
(cf.\ formula (\ref{h22formula})) which implies $\chi < 0 \iff h^{1,2} > 8 + h^{1,1} + h^{1,3}$, together with the small range of values of $h^{1,2}$ compared to those of $h^{1,1}$ and $h^{1,3}$.
Structures with a band-like appearance as in \cref{fig:h22-hodge-2,fig:chi-hodge-2,fig:h22-ws-2,fig:chi-ws-2} are also related to (\ref{h22formula}) and (\ref{chi}): assuming that both $h^{1,2}$ and $h^{1,1} + h^{1,3}$ have a preference for being even, $\chi/6 = 8 + h^{1,1} - h^{1,2} + h^{1,3}$ and $h^{2,2}/2=22+2h^{1,1}-h^{1,2}+2h^{1,3}$ will also tend to be even rather than odd; in Figs.~\ref{fig:chi-hodge-2} and \ref{fig:chi-ws-2} we can even see a preference for $\chi/6$ to be a multiple of 4.

While our main focus here is on the weight systems that determine reflexive polytopes, one should not ignore the ones giving polytopes with the IP property that lack reflexivity.
On the one hand they are indispensable ingredients in a full classification of reflexive polytopes.
On the other hand polytopes of this type may well be interesting on their own.
Originally reflexivity was singled out as the condition that leads
to smooth Calabi--Yau hypersurfaces in toric varieties of dimension up to four~\cite{alg-geom/9310003}.
If one does not insist on smoothness, which is not even guaranteed by reflexivity for polytope dimension $d>4$ anyway, the following setup becomes important.
In a notation in which $[\D]$ stands for $\mathrm{conv}(\D\cap M)$ (with an analogous definition for polytopes in $N_\IR$) a special role is played by IP polytopes that satisfy 
\beq \D = [\D] = [[\D^*]^*] \eeql{pseudo}
(the first condition $\D = [\D]$ just means that $\D$ is a lattice polytope).
Such polytopes are called \emph{almost reflexive} \cite{2011arXiv1103.2093M} or
\emph{pseudoreflexive} \cite{Batyrev:2017gny} and give rise to well-defined
singular varieties of Calabi--Yau type.

We will now argue that our polytopes $\D_\mathbf{q}$ satisfy condition
(\ref{pseudo}).
If we denote by $\nabla_\mathbf{q}$ the simplex in $N_\mathrm{coarsest}$
determined by some weight system $\mathbf{q}$ with the IP property, then
$\D_\mathbf{q} = [\nabla_\mathbf{q}^*] \subseteq \nabla_\mathbf{q}^*$,
hence $\D_\mathbf{q}^* \supseteq \nabla_\mathbf{q}$; since $\nabla_\mathbf{q}$ is
a lattice polytope this implies $[\D_\mathbf{q}^*] \supseteq \nabla_\mathbf{q}$
and therefore
$[[\D_\mathbf{q}^*]^*] \subseteq [\nabla_\mathbf{q}^*] = \D_\mathbf{q}$.
Conversely, $[\D_\mathbf{q}^*] \subseteq \D_\mathbf{q}^*$ gives 
$[\D_\mathbf{q}^*]^* \supseteq \D_\mathbf{q}$ which implies 
$[[\D_\mathbf{q}^*]^*] \supseteq \D_\mathbf{q}$ because $\D_\mathbf{q}$ is a
lattice polytope.
Therefore indeed $[[\D_\mathbf{q}^*]^*] = \D_\mathbf{q}$.
This fits well with the fact that both the lattice polytope $\nabla_\mathbf{q}$
and $[\nabla_\mathbf{q}^*]$ are IP polytopes, which means that $\nabla_\mathbf{q}$
is \emph{almost pseudoreflexive} in the sense
of Def.~3.6 and Prop.~3.4 of Ref.~\cite{Batyrev:2017gny}, whose Corollary 3.5
also implies that $\D_\mathrm{q}$ satisfies condition (\ref{pseudo}).\\

\paragraph{Acknowledgements:}
{
\uchyph=0
The authors thank Victor Batyrev for email correspondence and Roman Schönbichler for helpful discussions.
We are grateful to the Vienna Scientific Cluster for unbureaucratically providing computing time and in particular to Ernst Haunschmid for explanations on how to use these resources.
F.S.~has been supported by the Austrian Science Fund (FWF), projects P~27182-N27 and P~28751-N27.
}

\newpage
\section*{Appendix: Visualization of results}

\clearpage





\begin{figure}[H]
  \centering
  \includegraphics{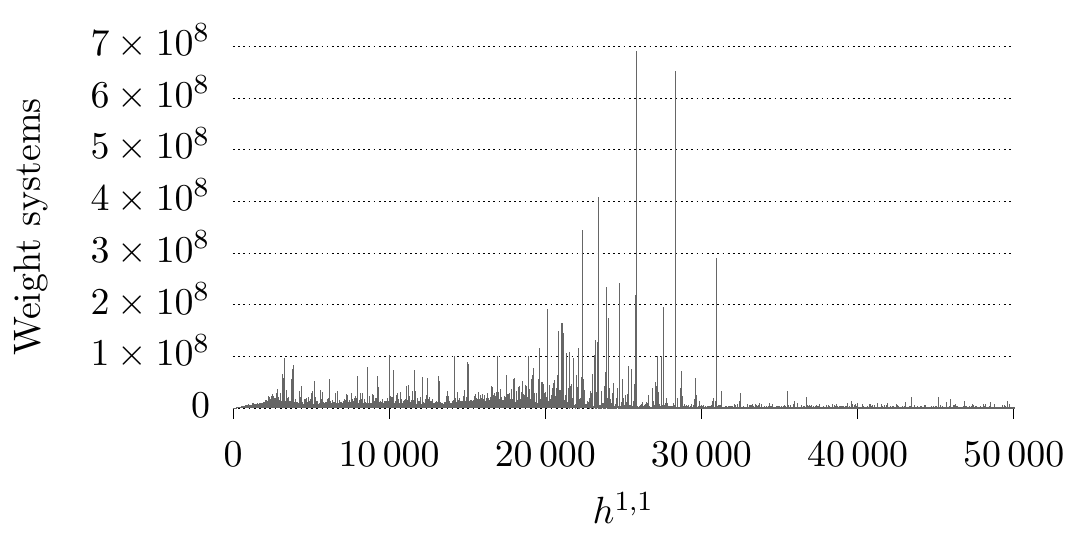}
  \caption{Numbers of weight systems leading to a given value of $h^{1,1}$}
  \label{fig:h11-ws-1}
\end{figure}
\begin{figure}[H]
  \centering
  \includegraphics{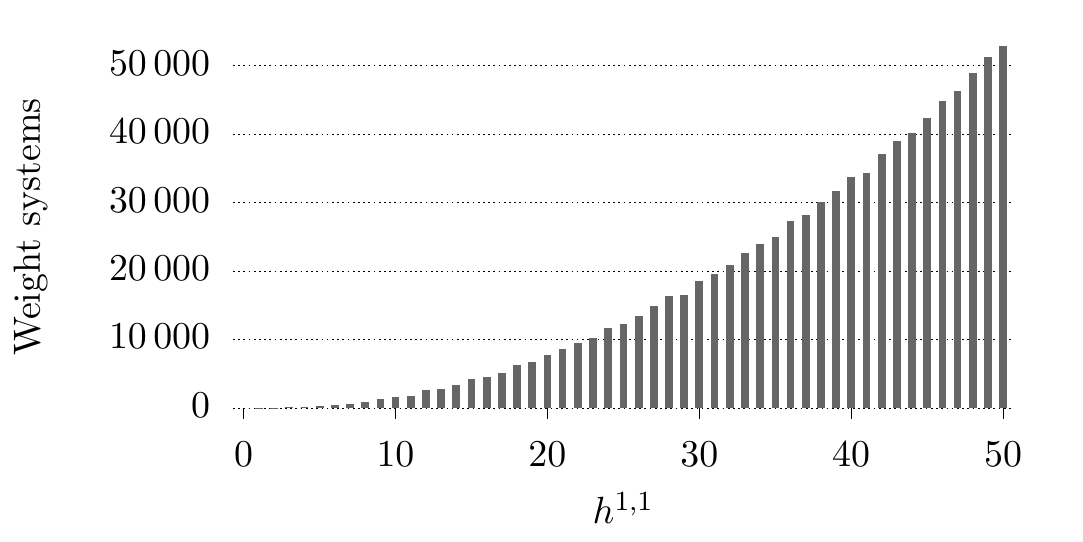}
  \caption{Numbers of weight systems leading to a given value of $h^{1,1}$}
  \label{fig:h11-ws-3}
\end{figure}
\begin{figure}[H]
  \centering
  \includegraphics{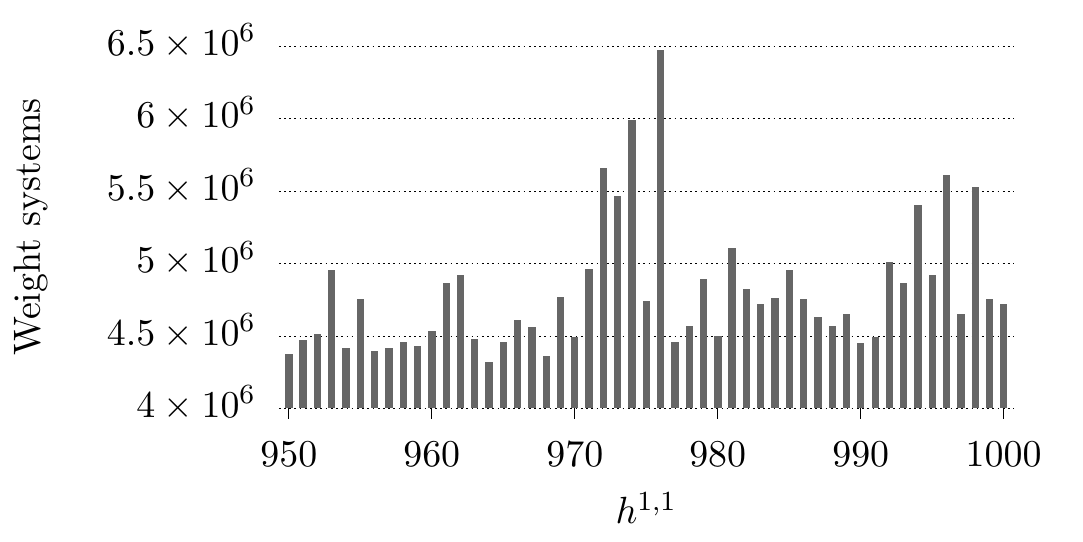}
  \caption{Numbers of weight systems leading to a given value of $h^{1,1}$}
  \label{fig:h11-ws-2}
\end{figure}

\begin{figure}[H]
  \centering
  \includegraphics{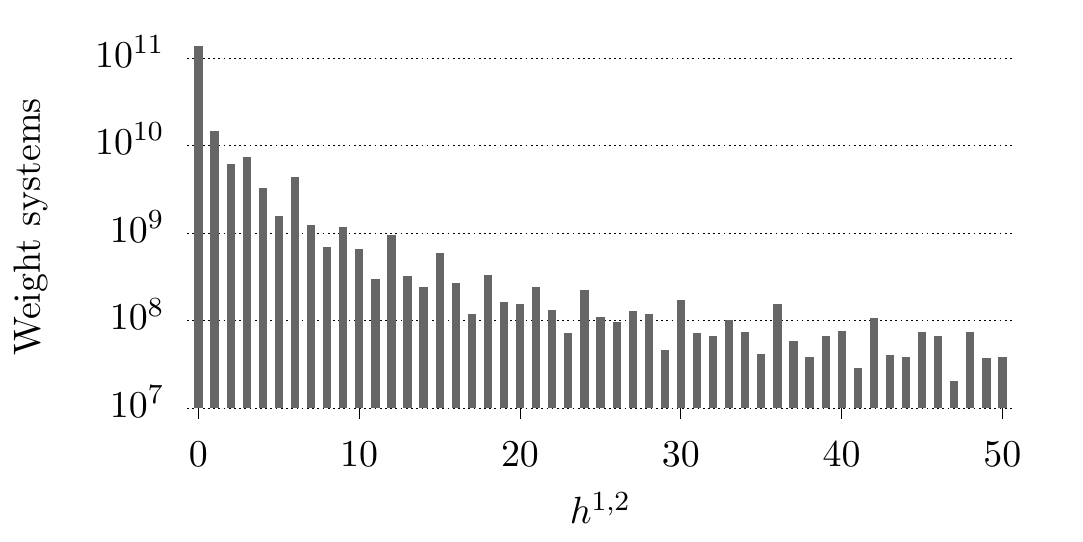}
  \caption{Numbers of weight systems leading to a given value of $h^{1,2}$}
  \label{fig:h12-ws-1}
\end{figure}
\begin{figure}[H]
  \centering
  \includegraphics{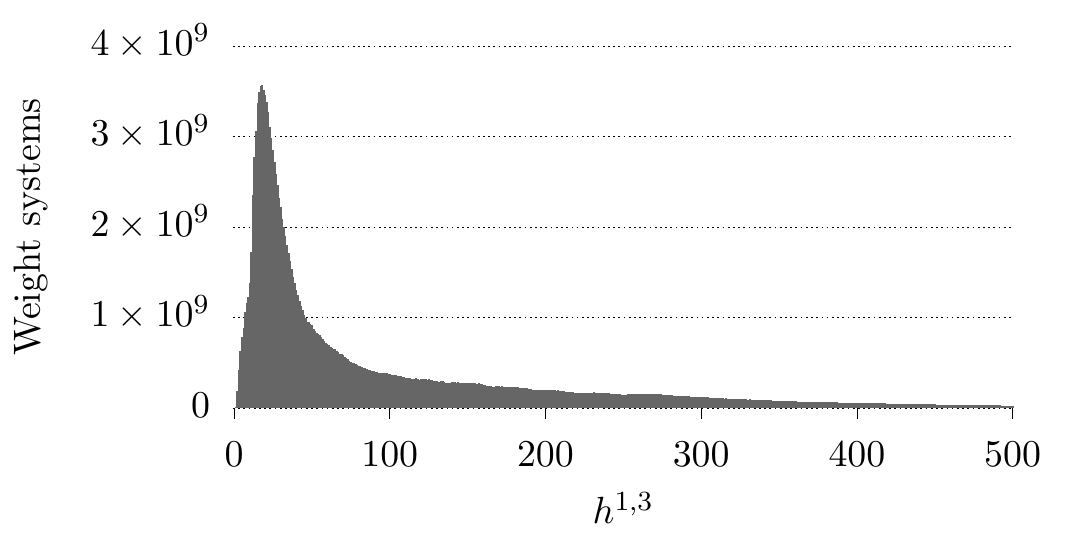}
  \caption{Numbers of weight systems leading to a given value of $h^{1,3}$}
  \label{fig:h13-ws-1}
\end{figure}
\begin{figure}[H]
  \centering
  \includegraphics{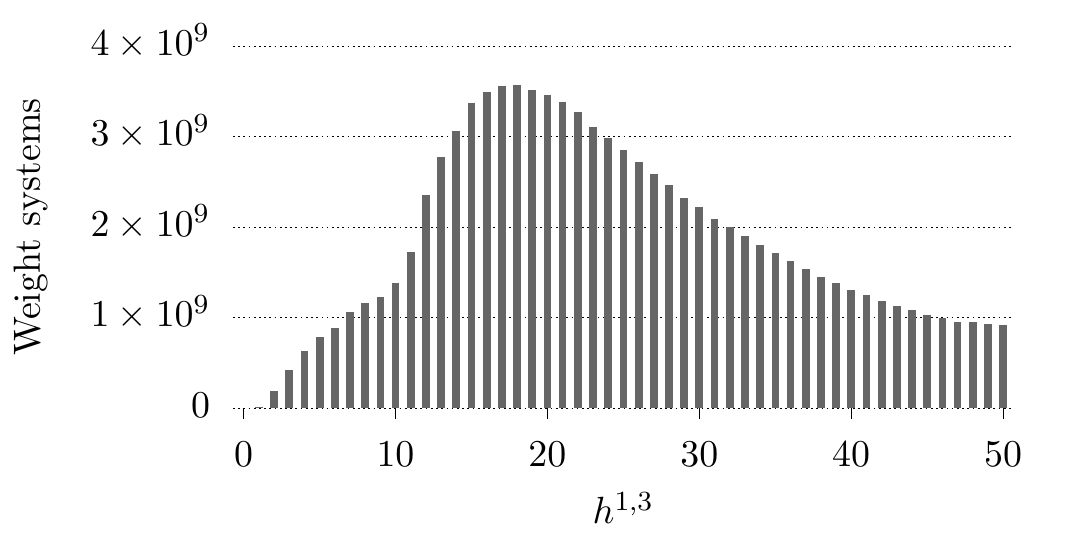}
  \caption{Numbers of weight systems leading to a given value of $h^{1,3}$}
  \label{fig:h13-ws-2}
\end{figure}

\begin{figure}[H]
  \centering
  \includegraphics{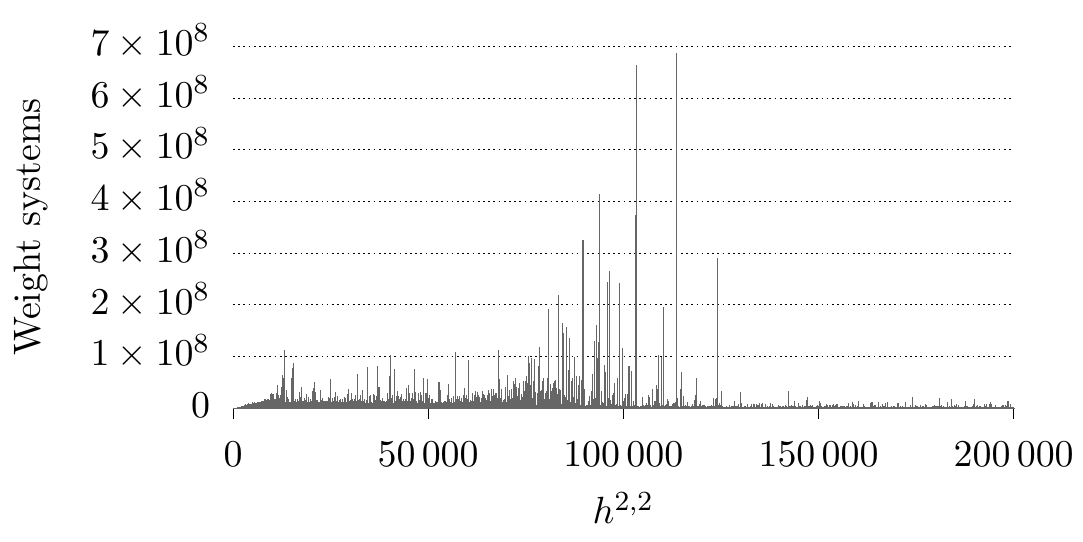}
  \caption{Numbers of weight systems leading to a given value of $h^{2,2}$}
  \label{fig:h22-ws-1}
\end{figure}
\begin{figure}[H]
  \centering
  \includegraphics{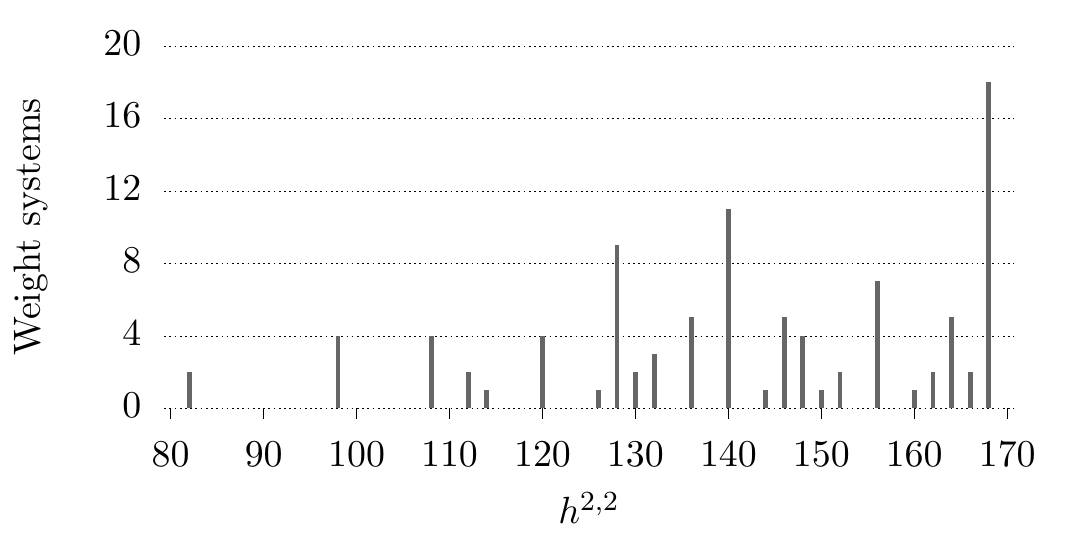}
  \caption{Numbers of weight systems leading to a given value of $h^{2,2}$}
  \label{fig:h22-ws-3}
\end{figure}

\begin{figure}[H]
  \centering
  \includegraphics{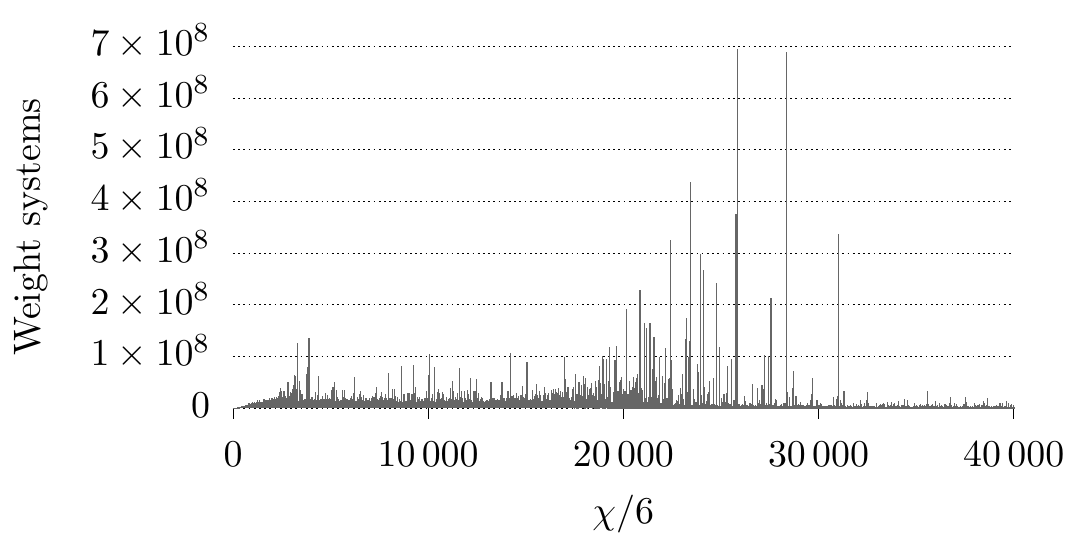}
  \caption{Numbers of weight systems leading to a given value of $\chi$}
  \label{fig:chi-ws-1}
\end{figure}

\bibliographystyle{utphys}
\bibliography{wff}

\end{document}